\documentclass{raa}            % referee version: for submission
\pdfoutput=1
\usepackage{graphicx,times}  %for PS/EPS graphics inclusion, new
\usepackage{natbib}
\usepackage{amssymb,amsmath}
\usepackage{amsmath}
\usepackage{multirow}
\usepackage{tabularx}
\usepackage{booktabs}
\usepackage[flushleft]{threeparttable}
\usepackage{lipsum}
\usepackage{enumitem}
\usepackage{hyperref}
\usepackage{siunitx}
\usepackage{rotating}
\usepackage{adjustbox} 
\bibpunct{(}{)}{;}{a}{}{,}

\usepackage{tikz,xcolor,hyperref}

\definecolor{lime}{HTML}{A6CE39}
\DeclareRobustCommand{\orcidicon}{%
	\begin{tikzpicture}
		\draw[lime, fill=lime] (0,0) 
		circle [radius=0.16] 
		node[white] {{\fontfamily{qag}\selectfont \tiny ID}};
		\draw[white, fill=white] (-0.0625,0.095) 
		circle [radius=0.007];
	\end{tikzpicture}
	\hspace{-2mm}
}

\foreach \x in {A, ..., Z}{%
	\expandafter\xdef\csname orcid\x\endcsname{\noexpand\href{https://orcid.org/\csname orcidauthor\x\endcsname}{\noexpand\orcidicon}}
}

\begin{document}

  \title{Main-sequence star super-flare frequency based on entire Kepler data
}
%   \subtitle{I. Place Your Subtitle Here}

   \volnopage{Vol.0 (20xx) No.0, 000--000}      %%preserved for Editor. DOn't remove!
   \setcounter{page}{1}          %%starting page, preserved for Editor. DOn't remove!

   \author{A. k. Althukair
      \inst{1,2}\orcidA{}
   \and D. Tsiklauri
      \inst{1}\orcidB{}
      }

   \institute{Department of Physics and Astronomy, School of Physical and Chemical Sciences, Queen Mary University of London,
   	Mile End Road, London, E1 4NS,
   	UK; {\it a.k.althukair@qmul.ac.uk}, {\it d.tsiklauri@qmul.ac.uk}\\
        \and
             Physics Department, College of Sciences, Princess Nourah Bint Abdulrahman University, Riyadh, PO Box 84428, Saudi Arabia\\
\vs\no
   {\small Received 20xx month day; accepted 20xx month day}}

\abstract{ We wrote and used an automated flare detection Python script to search for super-flares on main-sequence stars of types A, F, G, K, and M in Kepler's long-cadence data from Q0 to Q17. We studied the statistical properties of the occurrence rate of super-flares. For the G-type data set, we compared our results with the previous results of Okamoto et al. 2021 by splitting the data set into four rotational bands. We found similar power law indices for the flare frequency distribution. Hence, we show that inclusion of high-pass filter, sample biases, gyrochronology and completeness of flare detection is of no significance, as our results are similar to Okamoto et al. 2021. We estimated that a super-flare on G-type dwarfs of energy of $10^{35}$ erg occurs on a star once every 4360 years. We found 4637 super-flares on 1896 G-type dwarfs. Moreover, we found 321, 1125, 4538 and 5445 super-flares on 136, 522, 770 and 312 dwarfs of types A, F, K and M, respectively. We found that the occurrence rate ($dN/dE$) of super-flares versus flare energy, $E$, shows a power-law distribution with $dN/dE \propto E^{-\alpha}$, where $\alpha \simeq$ 2.0 to 2.1 for the spectral types from F-type to M-type stars. In contrast, the obtained $\alpha \simeq$ 1.3 for A-type stars suggests that the flare conditions differ from the rest spectral-type stars. We note an increase in flare incidence rate in F-type to M-type stars and a decrease in A-type to F-type stars.
}
   \authorrunning{Althukair \& Tsiklauri }            %author_head in even pages
   \titlerunning{Main-sequence star super-flare frequency }  % title_head in odd pages

   \maketitle
%% The author head (on even pages) and the title head (on odd pages) will be
%% automatically extracted from \author{} and \title{}. Whenever the title is too long,
%% you will be asked to supply a shorter one by inserting either \authorrunning{} or
%% \titlerunning{} before \maketitle. Anyway, you can specify your own heads.
%%
%%
%% Note: In the following text body of your manuscript, please note several differences from
%%       other major journals:
%% (1) \subsection{Please Capitalize the First Letter of Each Notional Word in Subsection Title}
%% (2) Please Capitalize the First Letter of Each Notional Word in all tables' captions

%
%________________________________________________ sections below
%
\section{Introduction}           %% first-level sections will be auto-capitalized
Flares are unpredictable events defined as a sudden, intense brightening caused by a large burst on the stellar surface as a result of magnetic energy release during the reconnection of twisted magnetic fields in the outer atmosphere of stars, which are typically located above or near star-spots \citep{Shibata_Magara2011,Walkowicz2011}. Flares produce electromagnetic radiation with a wide range of wavelengths, from long-wavelength radio waves to short-wavelength gamma rays \citet{Davenport_2016}.

In general the importance of Solar flares is due to their potential hazard to humankind,  particularly damages to space technology such as satellites and terrestrial  electrical power grids and pipelines via current surges generated by geomagnetic storms in the conductive layers of the Earth. \citet{Eastwood_2017} considers  an economic impact of space weather including the solar flares. \citet{Baumen_2014} made an estimate for a 1989 Quebec-like event, that the global economic impact would range from $2.4-3.4$ trillion dollars over a year.

However, the Sun is not the only star on which flares occur. Flares can arise in nearly all main sequence stars with exterior convective envelopes,  including both cool and hot stars \citet{Pettersen1989}, although they are more common in low-mass stars such as M dwarfs \citep{Walkowicz2011,Davenport_2016}. Stellar flares are thought to be generated by the same process by which solar flares occur through magnetic reconnection \citet{Davenport_2016}. X-ray observations show that young stars can produce "super-flares" \citet{Schaefer2000}, described as flares with radiative energy greater than $10^{33}$ erg. These young stars rotate quickly, with rotational periods of only a few days. Even in their quiescent phase, fast-rotating stars are strong X-ray sources, indicating a strong magnetic field and often showing large star spots \citep{Pevtsov2003,Cliver2022}. Thus, super-flares were thought impossible on the old, slow-rotating Sun. However, \citet{Schaefer2000} found nine super-flares with energies $10^{33\textendash3 8}$ erg in ordinary solar-type stars by analyzing previous astronomical data. White light flares are flares in the visible continuum \citet{Namekata2017}. White light emission from a solar flare was first seen in 1859 \citep{Carrington1859,Pitkin_2014,Namekata2017}. Several studies of white-light flares observation were carried out by \citep{Mathioudakis2003, Mathioudakis2006}

Due to the lack of spatial resolution on stars surfaces, the study of stellar flares is limited to photometry or spectroscopy \citet{Walkowicz2011}. Therefore, the Kepler mission made it possible to study stellar flares in detail \citet{Van-Doorsselaere2017}. Several studies have been conducted to analyse Kepler data and investigate stellar flares. The stellar rotation period, starspot area, and flare energy can all be estimated using Kepler photometric observations \citet{Cliver2022}.

\citet{Walkowicz2011} discovered 373 flaring stars on 23,000 cool dwarfs in the Kepler Quarter 1 long cadence data using a new flare measure, the photometric equivalent width ($\rm EW_{phot}$), which expresses the flare energy relative to the star's quiescent luminosity. Their findings suggest that M dwarfs flare more frequently but with shorter durations than K dwarfs and emit more energy. \citet{Pitkin_2014} provided a Bayesian method for identifying stellar flares in lightcurve data. The approach is based on the general premise that flares have a distinct form, in which there is a sudden increase with a half Gaussian shape accompanied by an exponential decline. During 120 days of Kepler observations, \citet{Maehara2012} reported 365 superflares on 148 solar-type stars (G-type main sequence stars). \citet{Shibayama_2013} studied statistics of stellar super-flares.
These authors discovered that for Sun-like stars (with surface temperature 5600-6000 K and slowly rotating with a period longer than 10 days), the occurrence rate of super-flares with an energy of $10^{34}-10^{35}$ erg is once in 800-5000 yr. \citet{Shibayama_2013} confirmed the previous results of \citet{Maehara2012} in that the occurrence rate ($dN/dE$) of super-flares versus flare energy $E$ shows a power-law distribution with $dN/dE \propto E^{-\alpha}$, where $\alpha \sim 2$. Such occurrence rate distribution versus flare energy is roughly similar to that for solar flares. Kepler data was used to analyse 4944 super-flares observed on 77 G-type stars by \citet{Wu2015}. They found that the power-law index $\gamma$ of the frequency distribution of flares, as a function of their energy, is $2.04 \pm 0.17$, consistent with previous studies. However, eight stars that flare frequently had $\gamma$ values ranging from $1.59 \pm 0.06$ to $2.11 \pm 0.19$, suggesting that these stars may have different energy release processes. Moreover, they found that stars with shorter rotation periods tend to have larger $\gamma$ values. \citet{Notsu2016} searched for super-flares on G-type main sequence stars and detected more than 1500 super-flares on 279 stars using long cadence data from Q0-Q6, and 187 super-flares on 23 stars using short cadence data from Q0-Q17. Their results show that the occurrence frequency of super-flares $(\rm dN/dE)$ as a function of flare energy $(\rm E)$ follows a power-low function with an index of -1.5. According to their findings, the frequency of super-flares depends on rotation period, with the frequency showing an increase as the rotation period decreases. \citet{yang2017} presented a study on 540 M dwarf stars that have exhibited flare events using Kepler long-cadence data. They examined the flare activity, normalized flare energy, chromosphere activity, and starspot characteristics of M dwarf stars. They identified three phases of flare activity related to rotation periods and noted a steep rise in flare activity near M4. In addition, they found a positive correlation between starspot size and flare activity, as well as the power-law relationship between flare energy and chromospheric activity. Using LAMOST DR5, Kepler, and K2 Missions, \citet{Lu2019} conducted a statistical study on M-type stars to investigate the relationship between chromospheric activity, flares, and magnetic activity in relation to rotation periods. They found that the flare frequency is consistent with chromospheric activity indicators and that the equivalent widths of H$\rm \alpha$ and Ca $\rm II$ H have a significant statistical correlation with the flare amplitude. In addition, they confirmed that magnetic activity and rotation period have an effect on flares. The study also determined thresholds for flare time frequency based on specific values of H$\rm \alpha$ equivalent width and rotation period. \citet{Yang2019} detected 162,262 flare events on 3,420 flaring stars among 200,000 Kepler targets using long cadence  mode (LC) data from data release 25 (DR25). \citet{Notsu2019} used Gaia-DR2 stellar radius estimates from \citet{Berger2018} and updated the parameters to study the statistical properties of Kepler solar-type super-flare stars first described in their previous studies \citep{Maehara2012, Maehara2015, Maehara2017, Notsu2013, Shibayama_2013}. Their findings indicate that, more than 40\% of the (279) solar-type (G-type main sequence) super-flare stars in \citet{Shibayama_2013} were classified as subgiant. Old, slowly-rotating Sun-like stars experience super-flares with energies $5\times 10^{34}$ erg once every 2000 to 3000 years, while young, fast-rotating stars experience super-flares with energies up to $10^{36}$ erg. In addition, the maximum super-flare energy gradually decreases as the rotation period increases. Moreover, the maximum area of starspots in the early stages of a star's life is independent of the rotation period. However, as the star ages and its rotation slows, the maximum area of starspots rapidly decreases at a certain $P_{rot}$ value. Since the flare energy can be explained by the magnetic energy stored around starspots, these two declining trends are consistent \citet{Notsu2019}. The most recent statistical analyses of super-flares on solar-type (G-type main-sequence) stars using all of the Kepler primary mission data and the Gaia Data Release 2 catalog have been reported by \citet{Okamoto2021}. They developed an improved version of the flare-detection method on their previous studies \citep{Maehara2012, Maehara2015, Shibayama_2013}, which involved the application of a high-pass filter to remove rotational variations caused by starspots. In addition, the sample biases on the frequency of super-flares were investigated, taking into consideration both gyrochronology and the completeness of the flare detection. They found 2341 super-flares on 265 solar-type stars and 26 on 15 Sun-like stars. It was estimated by \citet{Okamoto2021} that Sun-like stars with slow rotation could experience solar super-flares with energies of $10^{34}$ erg once every  $\sim$ 6000 years.

\citet{Davenport_2016} reported the first automated search for stellar flares using the entire Kepler data set of Data Release 24, including long and short cadence data. Approximately 3,144,487 light curves were analyzed for 207,617 distinct objects. \citet{Davenport_2016} identified 851,168 flares on 4,041 stars and revealed a strong correlation between flares and the evolution of stellar dynamos as stars age by comparing the amount of activity of the flare with stellar rotation and the Rossby number. Using thresholds for the intensity increase, the increase in the running difference and the flare duration \citet{Van-Doorsselaere2017} created a new technique for automated flare detection and applied it to Kepler's long-cadence data in quarter 15. Out of the 188,837 stars in the Kepler field of view during Q15, 16,850 flares have been found on 6662 of them.

The flare frequency distribution (FFD) is used to characterize flare energy and frequency, and it follows a power-law relation denoted by $dN/dE \propto E^{-\alpha}$ \citet{Dennis1985}. The $\alpha$ index constrains the magnetic activity of various stars \citet{Gao2022}. Flare completeness detection and precise energy calculation are crucial during fitting FFDs, as \citet{Gao2022} stated. They corrected the completeness of flare detection based on the data from Kepler and TESS, and reprocessed the light curve uniformly, correcting the detection efficiency for each star.
They improved the completeness and accuracy of the energy calculation in  flare detection by injection and recovery tests into each star's original light curve for each flare event.

Kepler data analysis was not only limited to manual and automated methods, several studies have used machine learning to analyze Kepler data. \citet{Vida2018} presented a machine-learning-based code for detecting and studying flares. The code was evaluated on two targets for Kepler and Kepler's second mission (K2) long and short cadence data, respectively. The detected flares for these two targets, as well as their energy, were found to be consistent with earlier findings. \citet{Breton2021} implemented a machine learning analysis pipeline to obtain rotation periods for Kepler targets. The algorithm was used on K and M main-sequence dwarfs studied in \citet{Santos2019}, the rotation periods of a sample of 21,707 stars were computed with an accuracy of 94.2\%. Machine learning techniques were applied by \citet{Ofman2022} on the TESS datasets to discover exoplanet candidates, by using Kepler data of verified exoplanets as a part of the algorithm training stage and validation. \citet{Vasilyev2022} developed a new method
for identifying the true flare sources using pixel-level data. It would be helpful for automated flare detection.

The physical process that adequately describes solar flares is magnetic reconnection. The latter is 
rapid change of connectivity of magnetic field lines, during which
magnetic energy is converted into thermal energy (heating)
and kinetic energy of plasma outflows. 
A general framework of solar flares is well accepted \citep{Masuda_1994,Shibata_1995} however questions  such as: 
(i) how frequently and (ii) under what conditions super-flares occur
still remain largely un-answered.

Our motivation is four-fold: 
\begin{enumerate}[label=(\roman*)]
	
	\item To study the statistics of stellar super-flares on main-sequence G-type stars based on Kepler data in quarters $0-17$, and compare our findings with those of \citet{Okamoto2021}.
	\item To examine the effect of not including (i) the high-pass filter and (ii) analysis of sample biases on the incidence rates of super-flare, (iii) gyrochronology and (iv) the completeness of the flare detection, considered by \citet{Okamoto2021} on our results.
	\item To investigate how the flare statistics changes on the time scale of 17 quarters. i.e. what is $\alpha$ power-law for data quarters $0-6$, $7-17$ and $0-17$.
	\item To study the statistics of stellar super-flares on main-sequence stars of other spectral types A, F, K and M.
	\item To provide a Python script which finds super-flares automatically and only minimal human-eye analysis is needed.
\end{enumerate}
Section \ref{section:Kepler-Data} presents the Kepler spacecraft and its data. Section \ref{section:Method} presents the method used including the targets selection, flares detection, rotation period determination and the flare energy estimation. 
Section \ref{section:results} provides the main results of this study. In particular we first produce results for $Q0-Q6$, $Q7-Q17$ and $Q0-Q17$. Then we present a comparison with \citet{Okamoto2021}. Finally, we present the result of other star spectral types.
Section \ref{section:conclusion} closes this work by providing our main conclusions.

\section{Kepler data }\label{section:Kepler-Data}
Kepler spacecraft was launched in 2009 by NASA to search for exoplanets using the transit photometry method \citet{koch2010kepler}. This spacecraft carried a photometer telescope with an aperture of a $0.95$ m and $105{\circ}^{2}$ field-of-view (FOV) and designed to stare fixedly at one patch of sky in the constellations of Cygnus, Lyra and Draco, monitoring roughly 200000 stars continuously, to detect changes in brightness caused by planets passing in front of the stellar disc. \citep{Shibayama_2013,Davenport_2016,Yang2019}. The Kepler mission had two stages during its lifetime. Kepler's primary mission (K1) which lasted for four years from 2009 to 2013  when the spacecraft lost two wheels of the four reaction wheels on board. Thus, Kepler's second mission (K2) started in 2014 and carried out until 2018. Due to the problem with the telescope's reaction wheels, it observed around the ecliptic plane. Targets were observed by Kepler using two cadence modes. The long cadence  mode (LC) which provide one photometric data point with every 29.4 min, and the short cadence mode (SC) that provide one photometric data point with every 1 min \citet{Pitkin_2014}. In order to make the Kepler solar panels always  to face the Sun, Kepler rotates every 90 days. Thus Kepler's data divided into almost 90-day quarters from quarter 0 to 17 as an abbreviation (Q0 - Q17) except for  Q0, Q1, Q8 and Q17, which covers (9, 33, 67 and 32 days respectively). The observational period with the start and end date of each quarter are shown in Table\ref{tab:Kepler Data}.\\ 
Since the luminosity loss caused by planetary transits is usually less than one hundredth of the star's total brightness, Kepler is intended to obtain high-precision and long-period light curves of many stars \citet{Shibayama_2013}. As a result, Kepler is considered as a perfect platform to analyse stellar flares due to the significant sample size, the duration of the light curves and the photometric accuracy \citet{Davenport_2016}.
\begin{table}
	\centering
	\caption{The observational period {\rm $T$} and the start and end dates of each quarter.}
	\label{tab:Kepler Data}
	\begin{tabular}{cccc}
		\hline
		\hline
		{\rm Quarter} & {\rm $T$} & {\rm Strat Date} & {\rm End Date}\\
		& (days) &(UT) &(UT) \\
		\hline
		0 & 9 & 2009-05-02 & 2009-05-11\\
		1 & 33 & 2009-05-13 & 2009-06-15\\
		2 & 88 & 2009-06-20 & 2009-09-16\\
		3 & 89 & 2009-09-18 & 2009-12-16\\
		4 & 90 & 2009-12-19 & 2010-03-19\\
		5 & 95 & 2010-03-20 & 2010-06-23\\
		6 & 90 & 2010-06-24 & 2010-09-22\\
		7 & 90 & 2010-09-23 & 2010-12-22\\
		8 & 67 & 2011-01-06 & 2011-03-14\\
		9 & 97 & 2011-03-21 & 2011-06-26\\
		10 & 93 & 2011-06-27 & 2011-09-28\\
		11 & 97 & 2011-09-29 & 2012-01-04\\
		12 & 83 & 2012-01-05 & 2012-03-28\\
		13 & 90 & 2012-03-29 & 2012-06-27\\
		14 & 97 & 2012-06-28 & 2012-10-03\\
		15 & 98 & 2012-10-05 & 2013-01-11\\
		16 & 86 & 2013-01-12 & 2013-04-08\\
		17 & 32 & 2013-04-09 & 2013-05-11\\
		\hline
	\end{tabular}	
\end{table}

\section{The Method }\label{section:Method}

\subsection{Targets Selection}
We carried out an automated search for super-flares on main-sequence stars type (A, F, G, K, M) based on entire Kepler data, using our bespoke Python script on long cadence data of Data Release 25 (DR 25). The script can be found at (\citet{AT2023AFD},
AFD.py, v1.0.0, Zenodo, \href{https://zenodo.org/badge/latestdoi/578168613}{doi:10.5281/zenodo.7755912}, as developed on \href{https://github.com/akthukair/AFD}{GitHub}).
The parameters for all targets observed by Kepler have been taken from The NASA Exoplanet Archive. The algorithm we used based on the method of \citep{Maehara2012,Shibayama_2013}. All Kepler light curve data (2.5 TBytes) were obtained as a fits files from the Mikulski Archive for Space Telescope (MAST)
with kind assistance of  Deborah Kenny of STScI. Only long cadence targets were selected (with time resolution of 29.4 min). 
Since Kepler's optical aperture has a radius of 4-7 pixels \citet{Bryson2010}, and the pixel size of the CCDs is about four arcs \citet {van_2009}, it is thus quite possible that some targets are very close to each other on CCDs, which indicates that nearby star's brightness variations may influence the target star's flux \citep{Shibayama_2013,yang2017,Yang2019}. Due to this reason we calculated the angular distance between every two stars in the entire sample, which is about 200,000 stars and excluded pairs of neighbouring stars within 12 arcsec from the analysis, to avoid detecting fake flares on the target \citet{Shibayama_2013}. The overall number of samples that were excluded from the study according to this condition are about 6\%. The angular distance $\theta$ between two stars was calculated by the following equation:
\begin{equation}
	\theta =\cos^{-1} [ \sin\delta_1\sin\delta_2+\cos\delta_1\cos\delta_2\cos(\alpha_1-\alpha_2)],
\end{equation}
where $\alpha_1$ , $\alpha_2$ , $\delta_1$ , $\delta_2$ are the right ascensions and the declinations of the two stars in degree respectively. We used the Harvard Spectral classification to obtain the spectral type for each target as shown in Table \ref{tab:spectral-type}. Whereas the effective temperature and radius of main-sequence stars in different spectral type are as follows: $2400\leq T_{\rm eff}<3700$, and radius $\leq 0.7\ R_\odot$ for M-type, $3700\leq T_{\rm eff}<5200$ and radius of $0.7-0.96\ R_\odot$ for K-type, for G-type we used the same effective temperature of \citep{Maehara2012,Shibayama_2013} which range between $5100\leq T_{\rm eff}<6000 \ K$ and radius of $0.9-1.15\ R_\odot$, F-type has $6000\leq T_{\rm eff}<7500\ K$ and radius range between $1.15-1.4\ R_\odot$, and A-type has $7500\leq T_{\rm eff}<10,000\ K$ and radius between $1.4-1.8\ R_\odot$ for A-type. Due to the small number of A-type main-sequence stars that fall under these conditions, which makes the statistics inaccurate, we have not implemented radius restrictions for this spectral type. The total number of main-sequence stars  is 2222, 10307, 25442, 10898, 2653 for M-, K-, G-, F-, and A-type, respectively.
\begin{table}
	\centering
	\caption{The effective temperature $T_{\rm  eff}$, radius and number of stars for each spectral class.}
	\label{tab:spectral-type}
	\begin{tabular}{cccc}
		\hline
		\hline
		{\rm Class} & $T_{\rm  eff}$ & {\rm Radius} & $N_{\rm star}$\\
		& $(\rm K)$ &($\rm R_\odot$) & \\
		\hline
		\rm A & 7500 - 10000 & 1.4 - 1.8 &2653 \\
		\rm F & 6000 - 7500 & 1.15 - 1.4 & 10898\\
		\rm G & 5100 - 6000 & 0.9 - 1.15 &25442 \\
		\rm K & 3700 - 5200 & 0.7 - 0.96 &10307 \\
		\rm M & 2400 - 3700 & $\leqslant$ 0.7 &2222 \\		
		\hline
	\end{tabular}	
\end{table}
\subsection{Flare Detection Method}
\begin{figure}
	\centering
	\includegraphics[width=0.6\textwidth]{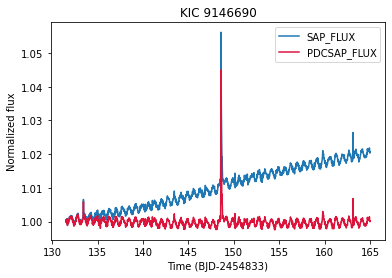}
	\caption{Light curve of KIC 9146690 using two types of flux. The simple aperture photometry flux SAP FLUX (blue) and the pre-searched conditioning SAP flux PDCSAP FLUX (red) which detrends the time variation.}
	\label{types_of_flux}
\end{figure}

Kepler light curves contain two kinds of flux, the Simple Aperture Photometry flux (SAP) and the Pre\textendash searched Conditioning SAP flux (PDCSAP), which has long term trends removed \citet{Davenport_2016}. Figure \ref{types_of_flux} illustrates the difference between these two types of fluxes. All light curves were analyzed using an algorithm with a similar technique to \citep{Maehara2012,Shibayama_2013}.

\begin{figure}[t]
	\centering
 \begin{subfigure}{\linewidth}
	\mbox{{\includegraphics[width=0.33\textwidth]{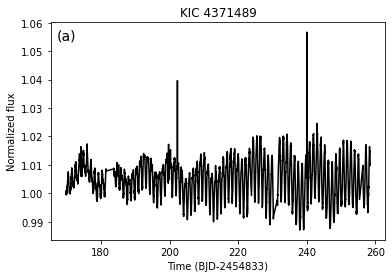}}\quad
		{\includegraphics[width=0.33\textwidth]{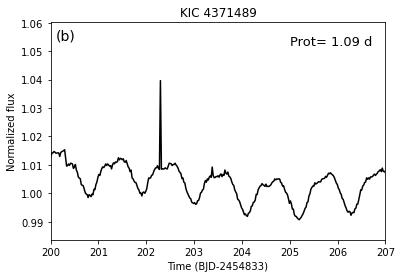} }\quad
		{\includegraphics[width=0.33\textwidth]{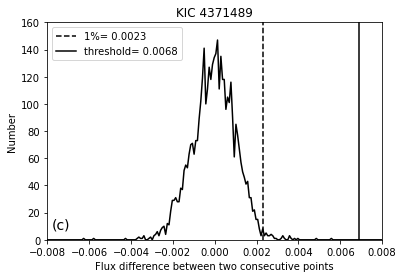} }}
	\mbox{{\includegraphics[width=0.33\columnwidth]{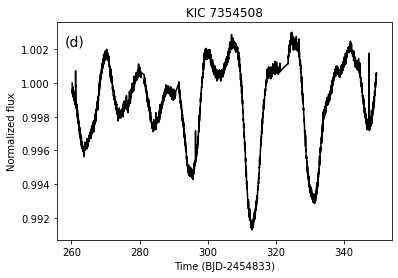}}\quad
		{\includegraphics[width=0.33\columnwidth]{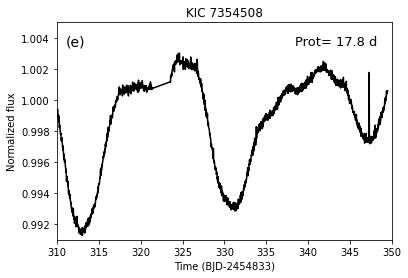} }\quad
		{\includegraphics[width=0.33\columnwidth]{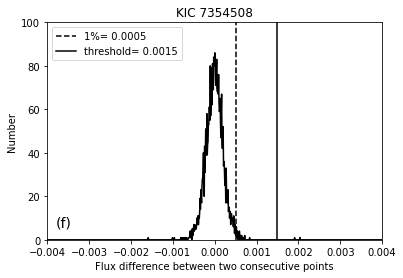} }}
 \end{subfigure}	
	\caption{Illustrations of flares detection method used by \citet{Shibayama_2013} in both fast and slowly rotating stars. (a) and (d) Show the light curve of KIC 4371489 and KIC 7354508 respectively. (b) A zoom-in into the light curve of KIC 4371489, showing a rotation period of 1.09 days. (e) A zoom-in into the light curve of KIC 7354508, showing a rotation period of 17.8 days. (c) and (f) show the distributions of brightness variation between every two neighbouring points of all the data points in the light curves of KIC 4371489 and KIC 7354508 respectively. The dashed vertical lines denote the value of 1\% of the total area under the curve, and the solid vertical lines denote the flare detection threshold.}
	\label{threshold}
\end{figure}

A brief description of this method is as follows. In order to be statistically accurate, after generating light curves of all selected stars using the PDCSAP flux,  we computed the distributions of brightness variation by calculating the flux difference in adjacent time intervals between every two neighbouring points of all the data points in the light curve. The purpose of this step, as clarified by \citet{Shibayama_2013}, is to avoid false flare detection and misdetection of short stellar brightness variation and not to overpass large flares. Then, we find such value of flux difference where the area under the distribution is equal to 1\% of the entire area. This value indicates a large flux difference between two adjacent points. To enhance the the threshold, we multiplied the 1\% value of the area by a factor of three. This threshold value has been chosen by \citet{Shibayama_2013} based on multiple tests. Examples of some of the outputs of this method, which give the distribution of brightness variations and the threshold determination, are shown in Figure \ref{threshold}. Figures \ref{threshold}(a) and \ref{threshold}(d) show light curves of KIC 4371489 at Q2 with a rotation period of 1.09 days and KIC 7354508 at Q3 with a rotation period of 17.8 days, respectively. Figures \ref{threshold}(b) and \ref{threshold}(e) show a zoom-in of these light curves, illustrating the rotational periods. Figures \ref{threshold}(c) and \ref{threshold}(f) show the distribution of brightness difference between every two neighbouring points of all the data points in the light curves of KIC 4371489 and KIC 7354508, respectively. The dashed vertical lines indicate 1\% of the total area under the distribution curves. The solid vertical lines indicate the threshold values of flares detection, which is equal to three times 1\% of the area under the curve. According to \citet{Shibayama_2013}, the detection threshold depends on the star's rotational period, and its brightness variation amplitude. For short-rotation-period stars (e.g., KIC 4371489), the distribution of brightness variations appears to extend larger than that for long-rotation-period stars (e.g., KIC 7354508). This large extension is because the difference in brightness between two successive data points is greater in stars with a short rotation periods than in stars with a long rotation periods, resulting in a greater value of flare detection threshold. Also, thresholds in stars with large brightness variation amplitudes are larger than those in stars with small brightness variation amplitudes. We defined the start time of a flare as the time when the flux difference of two consecutive points exceeds the threshold for the first time. 

\begin{figure}
	\centering
	\begin{subfigure}{\linewidth}
		\mbox{{\includegraphics[width=0.33\textwidth]{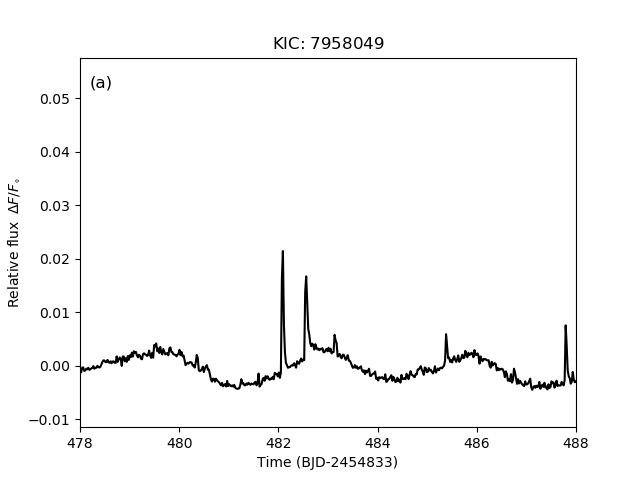}}\quad
		{\includegraphics[width=0.33\textwidth]{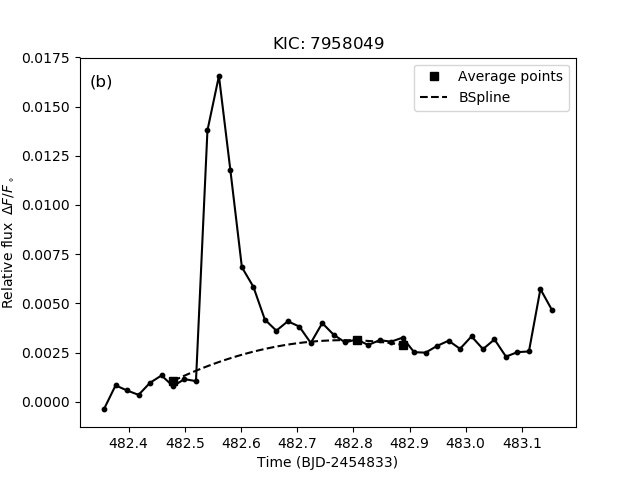}}\quad
		{\includegraphics[width=0.33\textwidth]{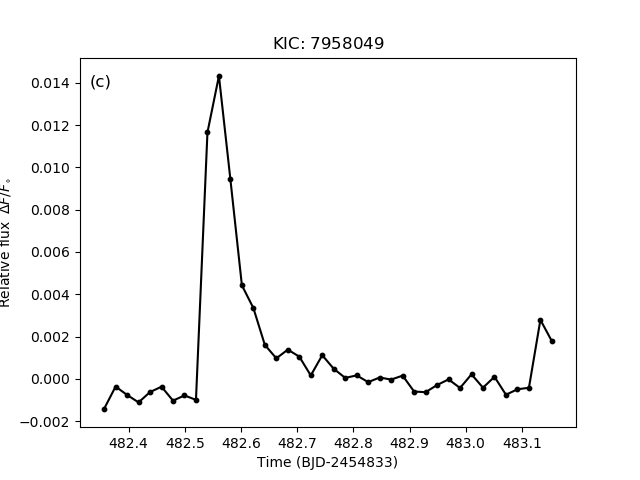} }}
	\end{subfigure}
		\caption{Demonstration of the removal of long-term brightness variations around the flare. (a) Part of KIC 7958049 light-curve around the flaring time (10 days of observation). (b) Fitting a B-spline curve (dashed curve) through three points around the flare (squares). The first point is an average of five data points just before the flare, the second point is an average of five data points after 5 h of the peak, and the third point is an average of five data points  8 h after the peak. All three averages are shown as squares. Note that the flux is normalized by the flux just before the flare $F_{\circ}$. (c) The produced light-curve after subtracting the B-spline curve from the relative flux.} 
		\label{BSpline}	
\end{figure}

To determine the flare end time, we computed the three standard deviations ($ 3\sigma $) of the brightness variation distribution. We used the relative flux ($\Delta F / F_{\rm  avg}$)  as shown in Figure \ref{BSpline}, where $\Delta F= F_{\rm  norm}(t)-F_{\rm  avg}$, with $F_{\rm  norm}(t)$ being the normalized flux of the light curve, $F_{\rm  avg}$ is the normalized flux average, and fitted a B-spline curve through three points distributed around the flare. Each of these three points is an average of five data points distributed as follows, the first average point just before the flare, the second average point around 5 hours after the flare maximum and the third average point around 8 hours after the flare maximum, see Figure \ref{BSpline} (b). The purpose of curve fitting is to remove long-term brightness variations around the flare \citet{Shibayama_2013}. After subtracting the B-spline curve from the original relative flux as in Figure \ref{BSpline} (c), we define the end time of the flare as the time when the relative flux produced by the subtraction becomes less than the value of $ 3\sigma $ of the distribution for the first time. The flare amplitude is given by:
\begin{equation}
	A=\frac{F_{\rm  max}-F_{\rm  avg_{2}}}{F_{\rm  avg}},
\end{equation}
where $F_{\rm  max}$ is the normalized flux at the flare peak, $F_{\rm  avg_{2}}$ is the normalized flux average of two points distributed around the flare. The first is the average of five data points before the start of the flare and the second is the average of five data points after the end of the flare, $ F_{\rm  avg}$ is the normalized flux average.  

After selecting the start and end time of each flare, we applied conditions to all flare candidates. These conditions are as follows: the duration of the flare should be longer than 0.05 days, which equals 72 minutes, which means at least 3 data points, and the period of the decline phase of the flare should be longer than the period of the increase phase. Only flare events that satisfy these conditions were analysed \citet{Shibayama_2013}. After selecting all flares that met the conditions, we checked by eye the light curve for each flare and eliminated false flares. We have not tested the pixel level data of stars showing flares as done in \citet{Shibayama_2013} for the reason of
simplicity. As can be seen in our results this test omission does  not
alter overall findings about e.g. flare occurance rate which us found to be
similar to \citet{Shibayama_2013}.

\subsection{Rotational Period Determination}
\begin{figure}
	\centering
	\begin{subfigure}{\linewidth}
		\mbox{{\includegraphics[width=0.5\textwidth]{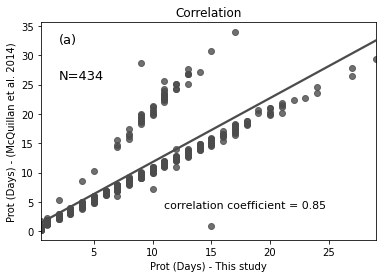}}\quad
		{\includegraphics[width=0.5\textwidth]{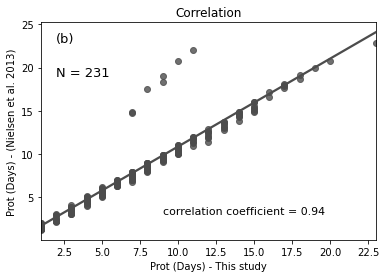} }}
	\end{subfigure}
		\caption{The comparisons of the periods determined in this study to those detected by \citet{McQuillan2014} and \citet{Nielsen2013} in (a) and (b), respectively, reveal a strong positive correlation in which both variables vary in the same direction. The x-axis shows the rotation periods found by this study, and the y-axis displays the rotational periods provided by \citet{McQuillan2014} and \citet{Nielsen2013}. N represents the number of stars in each comparison. The correlation coefficient between our periods and \citet{McQuillan2014} is 0.85, while it is 0.94 for \citet{Nielsen2013}}
		\label{correlation}
\end{figure}

We computed the brightness variation periods of light curves using the Lomb-Scargle periodogram, a common statistical method for identifying and describing periodic signals in unevenly sampled data \citet{VanderPlas2018}. We set the oversampling factor (number of samples per peak) in the periodogram to be five \citet{VanderPlas2018} and create a Lomb-Scargle periodogram for each light curve in each quarter from Q2 to Q16 using PDCSAP flux. The light curves data of Q0, Q1 and Q17 were excluded due to their short duration similarly to \citet{McQuillan2014}, as shown in Table \ref{tab:Kepler Data}. Further, we assigned the period corresponding to the highest power of the periodogram to be the rotation period for the Kepler ID in a certain quarter. To make the selection of star rotation period automatic rather than manual, we calculated this value with an accuracy of a day without the decimal part since fraction of a day would not significantly affect the results. For periods of less than a day, we set them to 0.5 days, and for periods of less than 0.1 days, we excluded them. According to \citet{McQuillan2014}, a good sign of actual astrophysical periodicity is that it can be found in different parts of the light curve. This is because peaks caused by systematics or artefacts are less likely to appear in multiple regions of the light curve. Therefore, in order to choose the accurate rotational period, we have selected the period that is most frequent in all quarters from Q2 to Q16 for each Kepler ID. We then required that the period chosen for all quarters should be identified in at least two distinct segments, following the \citet{McQuillan2014} technique, where the segment is defined as three consecutive Kepler quarters (Q2,Q3,Q4) - (Q5,Q6,Q7) - (Q8,Q9,Q10) - (Q11,Q12,Q13) - (Q14,Q15,Q16). It is worth noting that this method has only been applied to the 1897 Kepler IDs with super-flares that will be considered in \ref{0-17}. Using the segments technique, we determined the rotation periods of 548 flaring stars of the Kepler sample. In addition, we derived 222 rotation periods based on our most frequent period in all quarters. These selected periods showed significant correlation with those published in previous works. We used Excel's (CORREL) function to calculate the correlation coefficient. Where the correlation coefficient reaches about 0.99 for seven IDs in \citet{Yang_2019}, 0.99 for 70 IDs in \citet{Santos2021}, 0.97 for 54 IDs in \citet{McQuillan2014} and 0.95 for 54 IDs in \citet{Reinhold2015}, and 0.92 for 20 IDs in \citet{Nielsen2013}. After comparing the 770 rotation periods that we obtained from both approaches to those periods reported in \citet{McQuillan2014}, we found 434 common IDs with a correlation coefficient of 0.85 between their periods. Moreover, 231 common IDs in \citet{Nielsen2013} with a correlation coefficient of 0.94 between their rotation periods. Plots of these correlations are shown in Figure \ref{correlation}, where we used {\it regplot} function in Seaborn Python library to plot the linear regression model fit to the data. The x-axis in both figures represents the rotation periods obtained by this study. The y-axis represents the periods of rotation published by \citet{McQuillan2014} in \ref{correlation}(a) and \citet{Nielsen2013} in \ref{correlation}(b). In both figures, variables change in the same direction, indicating a significant positive correlation. The rotation periods of 80 stars were obtained from other works, including 67 from \citet{McQuillan2014}, 6 from \citet{Reinhold2015} and 7 from \citet{Santos2021}, while we found that 1,047 stars have no discernible rotation period. This could be due to three reasons as mentioned by \citet{yang2017}:
(i) the rotation period is longer than 90 days (a quarter), which makes it difficult (or impossible) to detect them in the star's frequency spectrum; (ii) at the accuracy level of Kepler, the light curve has a small amplitude due to the inclination angle and low activity level; (iii) fast-rotating stars have  spots in the poles \citet{schussler1992}, making detecting light variation through rotation hard.
The entire results on rotational period determination
can be found at \url{https://github.com/akthukair/AFD}.

\subsection{Flare Energy Estimation}
We determined the total energy of each flare from the star luminosity, flare amplitude and its duration, following \citep{Shibayama_2013,yang2017}.\\
The star luminosity $L_{\rm star}$, which is the total energy that a star produces in one second, depends on the radius of the star $R$ and the surface temperature $T_{\rm eff}$. This is given by the equation:
\begin{equation}
	L_{\rm star} = \sigma_{\rm  SB}T^4 _{\rm  eff} 4\pi R^2, 
\end{equation}
where $ \sigma_{\rm  SB}$ is the Stefan-Boltzmann constant, $4\pi R^2$ is the entire surface area of the star. \citet{Hawley1992,kretzschmar2011} found that the continuum emission released by white-light flare is compatible with blackbody radiation at about 9000 K. Therefore, in this study $T_{\rm  flare}$ assumed to be 9000 K according to \citep{Shibayama_2013,yang2017,gunther2020} and the luminosity for a blackbody emitting star is giving by:
\begin{equation} \label{Lflare}
	L_{\rm  flare}(t) = \sigma_{\rm  SB}T^4_{\rm  flare}A_{\rm  flare} ,
\end{equation}
where $A_{\rm  flare}$ is the area of the flare and can be estimated by the equation:
\begin{equation} \label{Aflare}
	A_{\rm  flare}(t) = C_{\rm  flare}(t) \pi R^2 \frac{\int R_{\lambda}B_{\lambda }(T_{\rm  eff}) d\lambda}
	{\int R_{\lambda}B_{\lambda }(T_{\rm  flare}) d\lambda} ,
\end{equation}
where $C_{\rm flare}$ is the flare amplitude for the relative 
flux, $R_{\lambda}$ is the response function of Kepler instrument \citet{caldwell2010kepler}. The photometer in Kepler use one broad bandpass range from 420 to 900 nm. $B_{\lambda}(T)$ is the Plank function at a given wavelength and it is given by:
\begin{equation}
	B_{\lambda}(T) = \frac{{2hc^2}/{\lambda^5}}{e^{{hc}/{\lambda k T}}-1},	
\end{equation} 
where $h$ is the Planck's constant, $c$ is the speed of light, $T$ is the temperature of the black body, $k$ is the Boltzmann's constant.\\
$L_{\rm  flare}$ can be calculated by substituting Eq.(\ref{Aflare}) into (\ref{Lflare}). Since $C_{\rm  flare}(t)$ is a function of time, $L_{\rm  flare}(t)$ is also a function of time. Therefore, the total energy of the flare is the integral of $L_{\rm  flare}$ over the flare duration, and is given by :
\begin{equation} \label{Eflare}
	E_{\rm  flare} =\int_{t_{\rm start}}^{t_{\rm end}} L_{\rm  flare}(t)dt .
\end{equation}

\section{The Results}\label{section:results}
This section presents the main results of this study.
\subsection{Super Flares on G-type Dwarfs}
\subsubsection{Super flares on G-type dwarfs in Q0-Q6}\label{subsubsection:0-6}
During 494 days of continuous observation of 25,440 G-type dwarfs, searching for super-flares using Kepler long cadence data, we found 1,298 super-flares on 588 G-type dwarfs. Among them, 229 super-flares on 132 slowly rotating stars. As for the Sun-like stars, which are known as stars with a surface temperature of $5600 K \leqslant T_{\rm  eff} < 6000 K$, a surface gravity of $\log \ g > 4.0$, and a rotational period exceeding ten days \citet{Shibayama_2013}, we found 151 super-flares on 93 Sun-like stars. The number of detected super-flares in this study is less by 16\% than the 1547 super-flares found in \citet{Shibayama_2013}. In contrast, the number of super-flare stars in this study is approximately 2 times the 279 super-flare stars  in \citet{Shibayama_2013}. Since interstellar activity varies, some stars have shown more than one super-flare, while others have shown only one. In comparison, we found that 161 stars out of 588 have more than one super-flare. While 427 stars showed only one super-flare. The number of Sun-like stars that exhibit more than one super-flare is 18 out of 93. While 75 Sun-like stars showed only one super-flare.
\par
\begin{figure}
	\centering
	\begin{subfigure}{\linewidth}
		\mbox{{\includegraphics[width=0.5\textwidth]{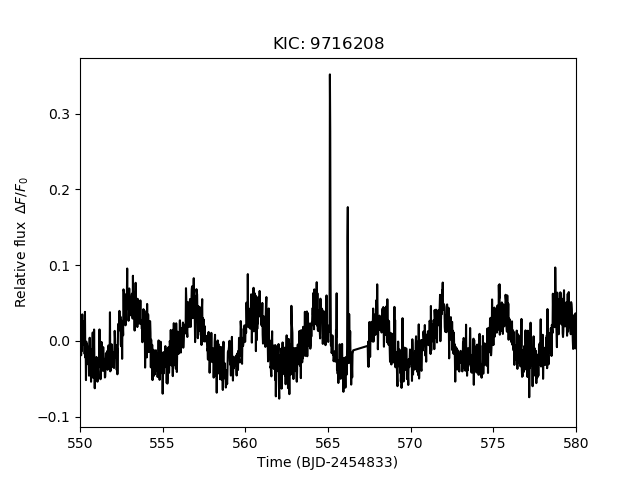}}\quad
		{\includegraphics[width=0.5\textwidth]{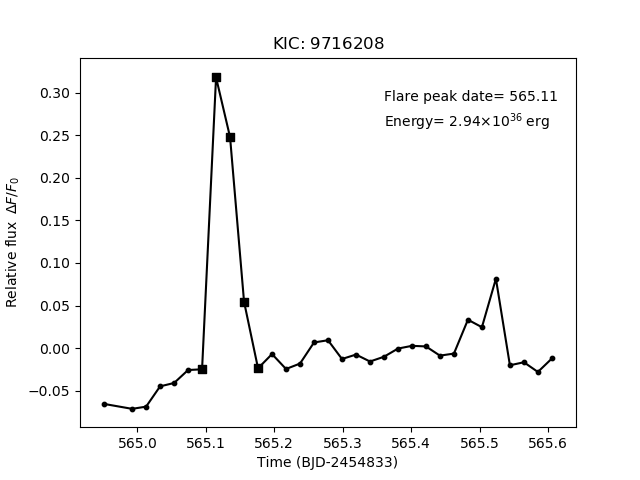} }}
		\mbox{{\includegraphics[width=0.5\textwidth]{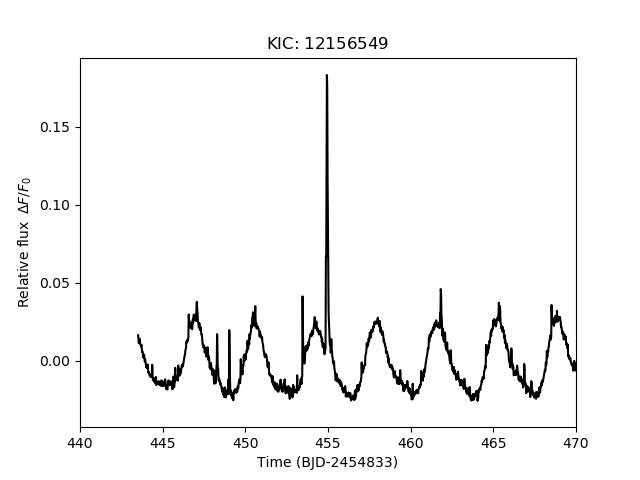}}\quad
		{\includegraphics[width=0.5\textwidth]{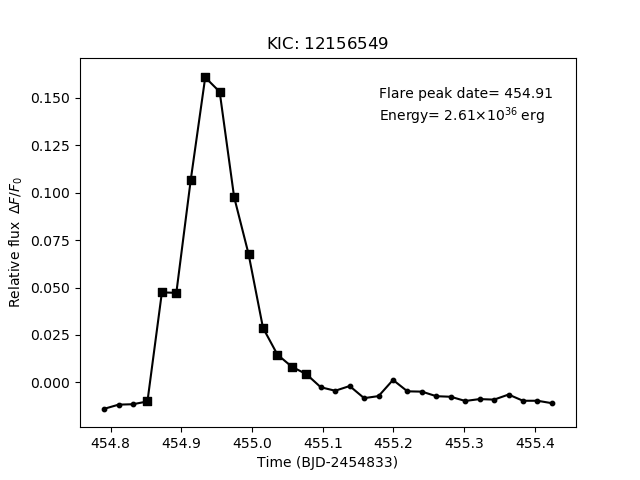} }}
		\mbox{{\includegraphics[width=0.5\textwidth]{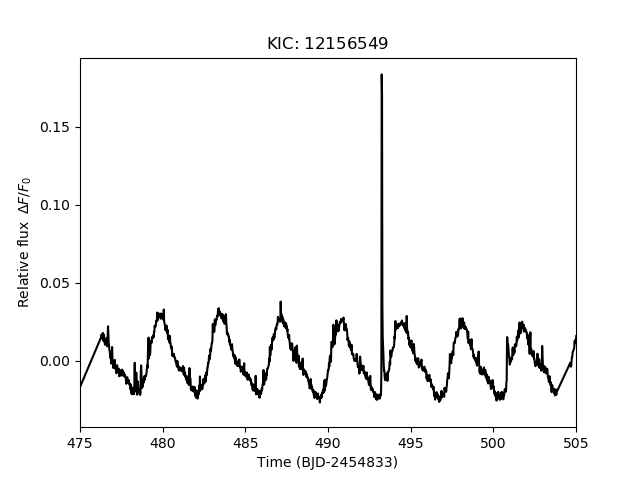}}\quad
		{\includegraphics[width=0.5\textwidth]{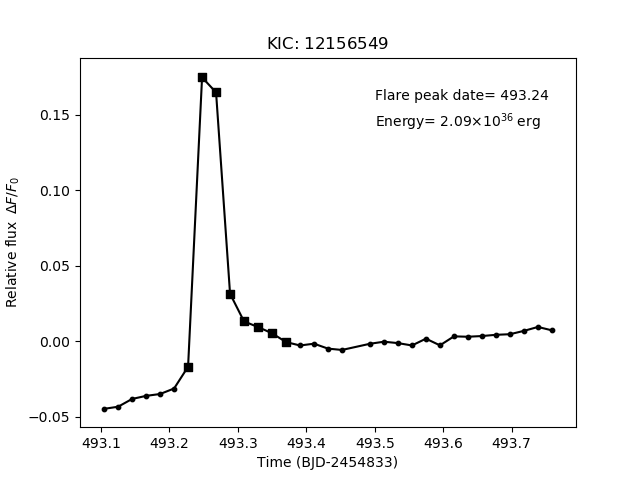} }}
		\mbox{{\includegraphics[width=0.5\textwidth]{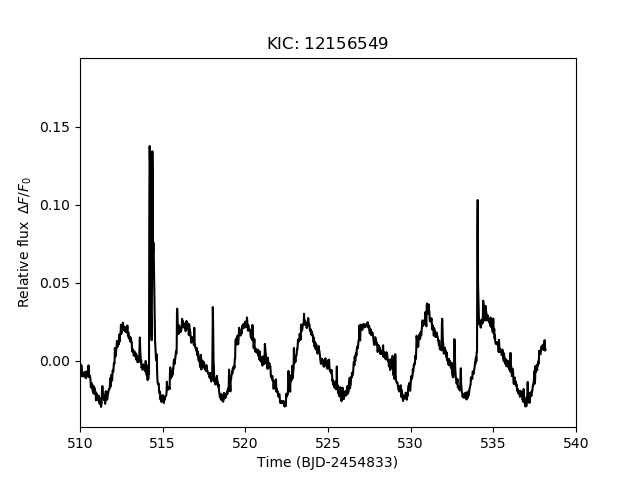}}\quad
		{\includegraphics[width=0.5\textwidth]{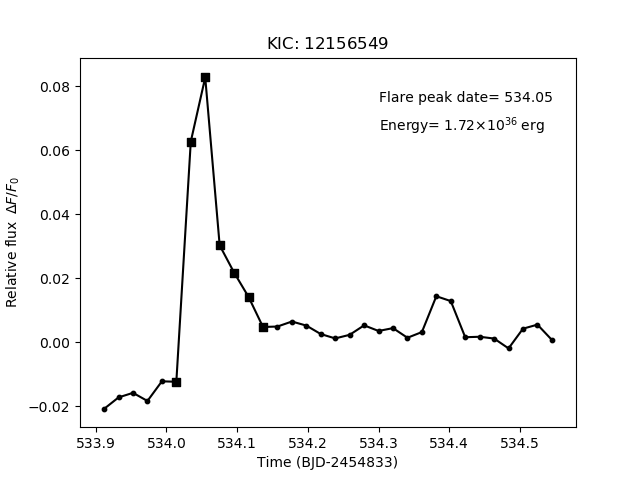} }}
	\end{subfigure}
		\caption{Four of the most energetic super-flares detected during quarter 0 to 6. The left panel shows the light curves of these super-flares over a period of 30 days. The super-flares that occurred on these light curves are enlarged in the right panels. The super-flare energy and peak date are displayed in the upper right corner. The black squares in the right panel correspond to the data points for the super-flares from beginning to end.} 
		\label{energy}	
\end{figure}

 The most energetic super-flare found had an energy of 
$2.94\times 10^{36}$ erg, an amplitude of 0.35 and lasted for about 0.08 days. Figure \ref{energy} shows four light curves with a 30-days observation period of the most energetic super-flares that we found  with their Kepler ID (left panels). The right panels display a zoom-in of these super-flares, showing their respective energy and their peak time. The black squares on the light curves indicate the data points of the super-flare from the time it starts until it ends.

\begin{figure}
	\centering
	\begin{subfigure}{\linewidth}
		\mbox{{\includegraphics[width=0.5\textwidth]{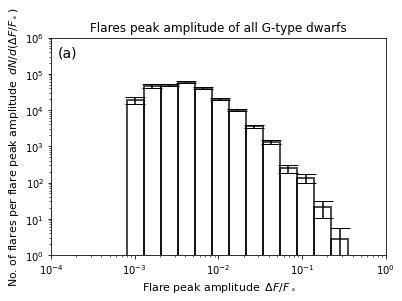}}\quad
		{\includegraphics[width=0.5\textwidth]{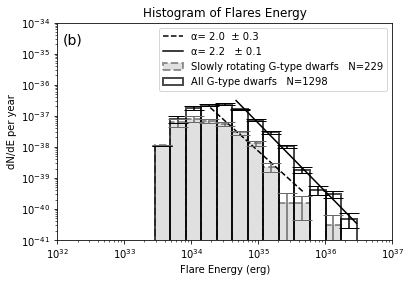} }}
		\mbox{{\includegraphics[width=0.5\textwidth]{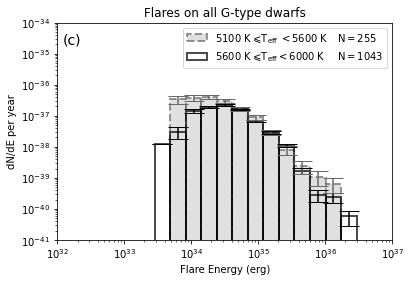}}\quad
		{\includegraphics[width=0.5\textwidth]{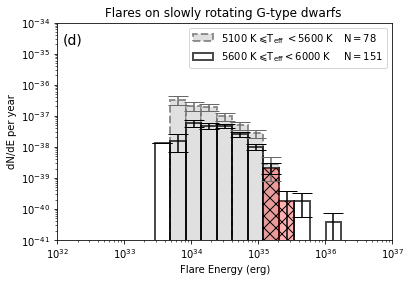} }}
		\mbox{{\includegraphics[width=0.5\textwidth]{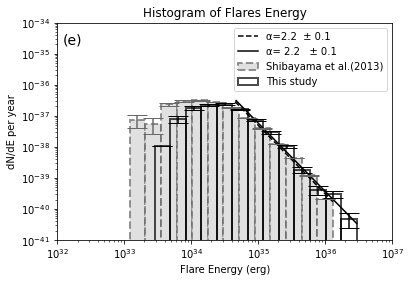}}\quad
		{\includegraphics[width=0.5\textwidth]{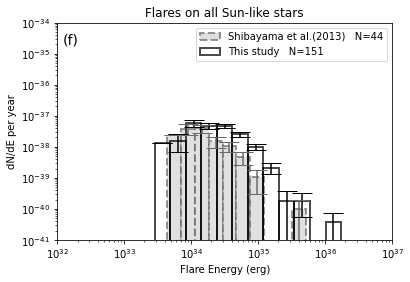} }}
	\end{subfigure}
		\caption{Log-log scale histograms showing the frequency distribution of super-flares on G-type dwarfs ($5100 K \leq T_{\rm  eff}<6000 K$, and $\log\ g>4.0$) in Q0-Q6. The error bar for each bin was calculated by taking the square root of the sum of the squared weights in that bin . (a) Distribution of the number of observed super-flares per recorded flare peak amplitude. (b) Distribution of flare frequency as a function of flare energy. The distributions of all stars and slowly rotating stars are shown by solid and dashed lines, respectively. All G-type dwarfs and slowly rotating G-type dwarfs have power-law indices of 2.2 and 2.0, respectively. (c) Distribution of flare frequency as a function of flare energy of all cool stars ($5100 K \leq T_{\rm  eff}<5600 K$, grey-dashed histogram), and all hot stars ($5600 K \leq T_{\rm  eff}<6000 K$ white-solid histogram) (d) The same as (c), but for flares on slowly rotating stars $P_{\rm  rot}>10$ days. The average energy of the two red bins were used to calculate the occurrence frequency of super flares with energy of $2.11\times10^{35}$ erg. (e) and (f) are the comparisons between the findings of this study and those of \citet{Shibayama_2013} on all G-type dawarfs and Sun-like stars, respectively.}
	\label{histogram-0-6}	
\end{figure}

In Figure \ref{histogram-0-6} are six log-log scale histograms demonstrating the frequency distribution of super-flares in quarters 0 to 6. We estimated the error bar for each bin using the equation: 
\begin{equation}
	{\rm err}=\sqrt{\sum w^2},	
\end{equation}
where $w$ represents the individual weights of the events that belong in that bin. As a result, when the number of the event in the bin is insufficient, the error bars in statistics are large \citet{Shibayama_2013}. 

Figure \ref{histogram-0-6}(a) represents the distribution of the number of observed super-flares per observed flare peak amplitude. The number of observed super-flares is 1298, and the measured amplitude range is approximately between $8\times10^{-4}$ to $3\times10^{-1}$. Figures \ref{histogram-0-6}(b,c and d) show the super-flares frequency distributions as a function of the flare energy. However, each histogram has a different collection of super-flare stars according to the purpose of the comparison. The x-axis indicates the energy of super-flares in erg, and the y-axis represents the number of super-flares per star per year per unit energy. Figure \ref{histogram-0-6}(b) show a comparison between frequency distributions of super-flares on all G-type dwarfs (white-solid histogram) and slowly rotating G-type dwarfs with $P_{\rm rot}>10$ days  (grey-dashed histogram). The number of detected super flares is 1298 for all G-type dwarfs and 229 for slowly rotating G-type dwarfs. Since the y-axis is the number of super-flares per star per year per unit energy, we determine the weight for each bin using:
\begin{equation}
	w =\frac{3.16\times10^{7}}{N_{\rm  os}\times D\times E}, 
\end{equation}
where $N_{\rm  os}$ is the number of observed stars, see table \ref{tab:Nstars}, $D$ is the duration of the observation period in seconds, and $E$ is the super-flare energy that belongs to that bin. Due to the lack of rotation period data, table \ref{tab:Nstars} was produced based on an estimate using table 6 from \citet{Shibayama_2013}, where we calculated the ratio of the number of cool, slowly-rotating stars and hot, slowly-rotating stars to the number of observed stars in \citet{Shibayama_2013}, and similarly for the fast-rotating stars. We then applied i.e. extrapolated these ratios to our data to estimate the number of fast and slowly rotating stars relative to their effective temperature. When comparing the distribution of super-flares on all G-type dwarfs and super-flares on slowly rotating G-type dwarfs, we can confirm that the occurrence frequency of super-flares on all G-type dwarfs is higher than the occurrence frequency of super-flares on slowly rotating G-type dwarfs. Furthermore, it is evident from the two fitted straight lines, (solid line) for all G-type dwarfs and (dashed line) for all slowly rotating G-type dwarfs, that the frequency distributions of super-flares on all G-type dwarfs and slowly rotating G-type dwarfs, follow a power-law relation given by: 
\begin{equation}
	\frac{dN}{dE}\propto E^{-\alpha},
\end{equation} 
where the index $\alpha \simeq 2.2 \ \pm 0.1$ for all G-type dwarfs and $\alpha \simeq 2.0 \ \pm 0.3$ for all slowly rotating G-type dwarfs. 
We note that this result is similar to that of \citet{Shibayama_2013}. Figure \ref{histogram-0-6}(c) show the distribution of super-flares as a function of energy on all cool G-type dwarf with $5100 K \leqslant T_{\rm eff} < 5600 K$ (grey-dashed histogram) with 255 observed super-flares, and all hot G-type dwarfs with $5600 K \leqslant T_{\rm eff} < 6000 K$ (white-solid histogram) with 1043 observed super-flares. Figure \ref{histogram-0-6}(d) the same as \ref{histogram-0-6}(c) but for slowly rotating G-type dwarfs with rotation period more than 10 days. The number of observed super-flares for cool slowly rotating G-type dwarfs is 78, and 151 for hot slowly rotating G-type dwarfs (Sun-like stars). From the two figures, we find that the frequency of super-flares is higher in cool G-type dwarfs than in hot G-type dwarfs, again confirming the previous results of \citep{Maehara2012,Shibayama_2013}. Note
that the difference in the occurrence rate of super-flares between the cool and hot stars in this study is not as large as in \citep{Maehara2012,Shibayama_2013}. We can explain this by noting that our sample's number of cool stars constitutes approximately 15.1\% of the total number of the sample while hot stars make up 84.9\% of the total number of the sample, see table \ref{tab:Nstars}. In contrast, in \citet{Shibayama_2013}, cool stars make up about 48.7\% of the total number of the sample, and hot stars constitute 51.3\% of the total number of the sample. In addition, if we compare the number of superstars relative to the observed stars, we find that this rate is higher in cool stars than in hot ones, see columns 5, 8 and 13 in table \ref{tab:Nflares}. 

\begin{table}
	\centering
	\caption{The number of observed G-type stars $N_{\rm  os}$ distributed according to their effective temperature $T_{\rm  eff}$ and rotational period $P_{\rm  rot}$.}
	\label{tab:Nstars}
	\begin{tabular}{cccc}
		\hline
		\hline
		$T_{\rm  eff}$  & $N_{\rm  os}$ & Slow  & Fast \\
		$K$ &  &$P_{\rm  rot}>10$\ d & $P_{\rm  rot}\leq10$\ d\\
		\hline
		5100-5600 & 3839 & 3518 &321 \\
		5600-6000 & 21603 & 19160 & 2443\\
		\hline
		Total & 25442 &22678 &2764 \\	
		\hline
	\end{tabular}	
\end{table}

We calculated the occurrence frequency rate of super-flares from the number of observed super-flares $N_{\rm  f}$, the number of observed stars $N_{\rm  os}$ and the observation duration $D$ \citep{Maehara2012,Shibayama_2013}. The super-flares energy distribution in Q0-Q6, which ranges from $2.84\times10^{33}$ erg to $2.94\times10^{36}$ erg, varies from Q7-Q17 and Q0-Q17, which ranges from $2\times10^{33}$ erg to $1.42\times10^{38}$ erg, resulting in a different energy bins distribution. Therefore, in order for the energy used to derive the occurrence frequency rate of super-flares to be close for each section of the analysis (Q0-Q6, Q7-Q17 and Q0-Q17), we used the average energy of the two red bins in \ref{histogram-0-6}(d), which equals $2.11\times10^{35}$ erg and is close to the energy $2.26\times10^{35}$ erg that used in Q7-Q17 and Q0-Q17, red bins in Figures \ref{histogram-7-17}(d) and \ref{histogram-0-17}(d). As a result, the rate of super-flares incidence with energy of $2.11\times10^{35}$ erg is $2.46\times10^{-4}$ flares per year per star, corresponding to a super-flare occurring on a star once every 4070 years. This result is 
within about 80\% of the occurrence rate of super-flares with energy of $10^{35}$ erg found by \citet{Shibayama_2013}, which equals to one super-flare in 5000 years for each star. Figures \ref{histogram-0-6}(e) and \ref{histogram-0-6}(f) show a comparison between the results of this study (solid-white-histogram) and \citet{Shibayama_2013} study (grey-dashed-histogram). Figure \ref{histogram-0-6}(e) depicts the frequency distribution of super-flares energy for all G-type dwarfs, $dN/dE \propto E^{-\alpha}$, where $\alpha \sim 2.2$  in each of the two studies. \ref{histogram-0-6}(f) show the same comparison but for Sun-like stars with 151 super-flares in this study and 44 super-flares in \citet{Shibayama_2013}. Note from the two comparisons that the estimated energy of super-flares in this study is higher than that of \citet{Shibayama_2013}. Our justification for this is related to the differences in the Kepler Data Release between DR 9 and DR 25 used in \citet{Shibayama_2013} and this study, respectively. Since the Kepler pipeline was updated between DR 9 and DR 25, PDC light curves for each target may also be updated. Also, the previous work may have used a slightly different parameters since the effective temperature and radius of some Kepler targets have been updated since then, resulting in different Kepler IDs for each spectral type. Another factor that may affect the energy value is that in this study, we used the flux difference rather than the flux to determine the start time of the flare. This can result in an extra data point between the flare's start and end times, resulting in higher energy.
To find how our method affects the energy calculation results, we run the script on 279 flare stars from \citet{Shibayama_2013} in order to compare the energy results drived form the script with those in \citet{Shibayama_2013}. We found 503 common super-flares with the same start-time of the flare between \citet{Shibayama_2013} and  our script result. Then we calculated correlation coefficient between the energy values and we observed an energy increase for our study, with a correlation coefficient of 0.66 between the two energies.
Such small correlation coefficient points to a sizeable difference 
the flare energy estimates between this study and \citet{Shibayama_2013}, for the reasons listed above.
\begin{table*}
	\centering
	\scriptsize
	\caption{The number of super-flares and super-flare stars on G-type dwarfs.}
	\label{tab:Nflares}
	\begin{tabular*}{\linewidth}{ @{\extracolsep{\fill}}ccccccccccccc}
		\hline
		\hline
		\multirow{3}{*}\centering{{\rm Quarters}} &
		\multirow{2}{*}\centering{{$T_{\rm eff}$}} &		
		\multicolumn{3}{c}{Slow} & 
		\multicolumn{3}{c}{Fast} &
		\multicolumn{2}{c}{Unknown $P_{\rm rot}$} &
		\multicolumn{3}{c}{Total} \\
		& & {$N_{\rm f}$} & {$N_{\rm fstar}$} &{${N_{\rm f}}/{N_{\rm os}}$}& {$N_{\rm f}$} & {$N_{\rm fstar}$} & {${N_{\rm f}}/{N_{\rm os}}$}&{$N_{\rm f}$} & {$N_{\rm fstar}$}&{$N_{\rm f}$} & {$N_{\rm fstar}$} & {${N_{\rm f}}/{N_{\rm os}}$} \\
		\hline
		$0-6$&$5100-5600$& 78 & 39 & 0.02 & 141 & 35 & 0.43  & 36 & 30 & 255 & 104 & 0.07 \\
		&$5600-6000$& 151 & 93 & 0.01 & 698 & 203 & 0.29 & 194 & 188 & 1043 & 484 & 0.05 \\
		\hline
		$7-17$&$5100-5600$& 195 & 88 & 0.06  & 303 & 59 & 0.94 & 144 & 116 & 642 & 263 & 0.17 \\
		&$5600-6000$& 288 & 180 & 0.015  & 1552 & 361 & 0.64 & 857 & 735 & 2697 & 1276 & 0.12 \\
		\hline
		$0-17$&$5100-5600$& 273 & 112 & 0.08  & 444 & 65  & 1.38 & 180 & 141 & 897 & 319 & 0.23\\
		&$5600-6000$& 439 & 243 & 0.023 & 2250 & 653 & 0.92& 1051 & 906 & 3740 & 1578 & 0.17  \\
		\hline		
	\end{tabular*}
		\begin{tablenotes}
		\small
		\item \textbf{Note:}\ $N_{\rm f}$ the number of super-flares, $N_{\rm fstar}$ the number of super-flare stars and ${N_{\rm f}}/{N_{\rm os}}$ the ratio of the number of super-flares to the number of observed stars ${N_{\rm os}}$. ${N_{\rm os}}$ can be found in table \ref{tab:Nstars}.
		\end{tablenotes}
\end{table*}

\subsubsection{Super flares on G-type dwarfs in Q7-Q17}
\begin{figure}
	\centering
	\begin{subfigure}{\linewidth}
		\mbox{{\includegraphics[width=0.5\textwidth]{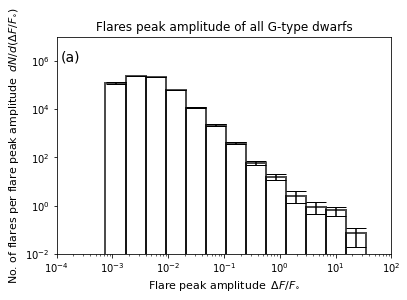}}\quad
		{\includegraphics[width=0.5\textwidth]{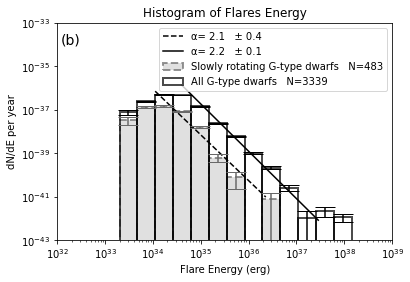} }}
		\mbox{{\includegraphics[width=0.5\textwidth]{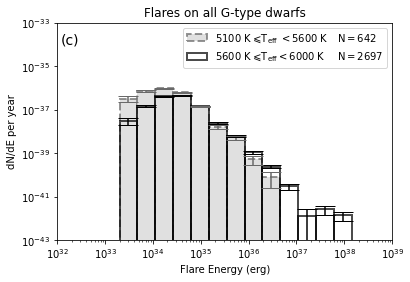}}\quad
		{\includegraphics[width=0.5\textwidth]{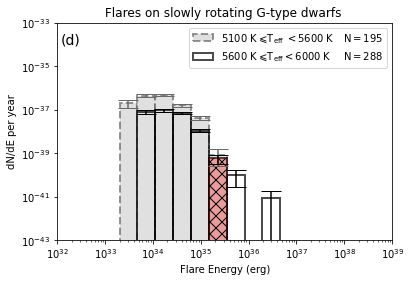} }} 
	\end{subfigure}
		\caption{The same as in Figure \ref{histogram-0-6} (a-d) but for Q7-Q17 data. The flare frequency distribution in (b) for all G-type dwarfs and slowly rotating G-type dwarfs have power-law indices of 2.2 and 2.1, respectively. The red bin in (d) was used to calculate the occurrence frequency of super flares with energy of $2.26\times10^{35}$ erg.}
		\label{histogram-7-17}	
\end{figure}

We detected 3339 super-flares on 1539 G-type dwarfs during the course of 930 days of continuous monitoring. 483 super-flares on 268 slowly rotating stars are among them. In the case of Sun-like stars, we found 288 super-flares on 180 of them. According to our findings, 318 G-type dwarfs out of 1539 exhibit several super-flares whereas 1221 show only one super-flare. A total of 38 out of 180 Sun-like stars have several super-flares, and 142 Sun-like stars have just one super-flare. The most powerful super-flare discovered during this period had an energy of $1.42\times10^{38}$ erg, an amplitude of 22.75, and a duration of 0.06 days. Figure \ref{histogram-7-17} depicts four log-log scale histograms displaying the frequency distribution of super-flares in quarters 7 to 17, similar to Figure \ref{histogram-0-6}, but without the comparisons in \ref{histogram-0-6}(e) and \ref{histogram-0-6}(f). The distribution of the number of observed super-flares per observed flare peak amplitude is shown in Figure \ref{histogram-7-17}(a). The total number of super-flares detected is 3339, with amplitudes ranging from $7.5\times10^{-4}$ to $35$. Figures \ref{histogram-7-17}(b,c and d) have similar format
as Figures \ref{histogram-0-6}(b,c and d). A comparison of super-flare frequency distributions on all G-type dwarfs (white-solid histogram) and slowly rotating G-type dwarfs (grey-dashed histogram) is shown in Figure \ref{histogram-7-17}(b). The power-law index for super-flare frequency distributions on all G-type dwarfs $\alpha \simeq 2.2 \ \pm 0.1$ and for slowly rotating G-type dwarfs $\alpha \simeq 2.1 \ \pm 0.4$. Figures \ref{histogram-7-17}(c) and \ref{histogram-7-17}(d) show the distribution of super-flares as a function of flare's energy according to the effective temperature of the star for all G-type dwarfs and slowly rotating G-type dwarfs respectively. where the number of detected super-flares is  642 super-flares on all cool G-type dwarf (grey-dashed histogram) and 2697 super-flares on all hot G-type dwarfs (white-solid histogram) in \ref{histogram-7-17}(c) and 195 super-flares on cool slowly rotating G-type dwarfs (grey-dashed histogram) and 288 super-flares on hot slowly rotating G-type dwarfs (white-solid histogram) in \ref{histogram-7-17}(d). The red bin in \ref{histogram-7-17}(d) shows that the occurrence frequency rate of super-flares with energy of $2.26\times10^{35}$ erg is $1.31\times10^{-4}$ flares per year per star, corresponding to a super-flare occurring on a star once every 7640 years.
We note that this value is somewhat larger that similar results
for Q0-Q6 and Q0-Q17.

\subsubsection{Super flares on G-type dwarfs in Q0-Q17}\label{0-17} 
\begin{figure}
	\centering
	\begin{subfigure}{\linewidth}
		\mbox{{\includegraphics[width=0.5\textwidth]{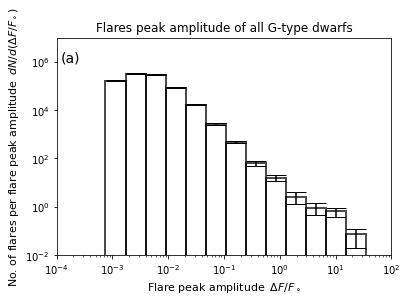}}\quad
		{\includegraphics[width=0.5\textwidth]{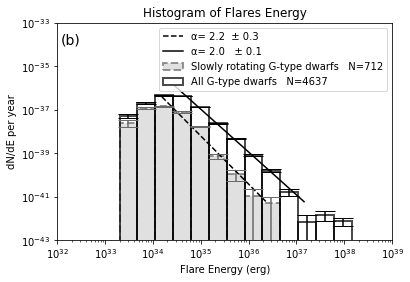} }}
		\mbox{{\includegraphics[width=0.5\textwidth]{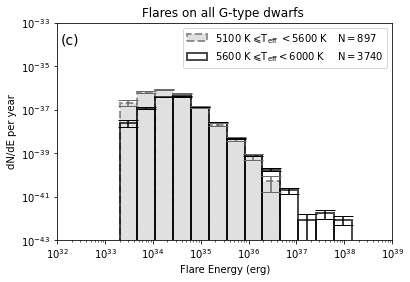}}\quad
		{\includegraphics[width=0.5\textwidth]{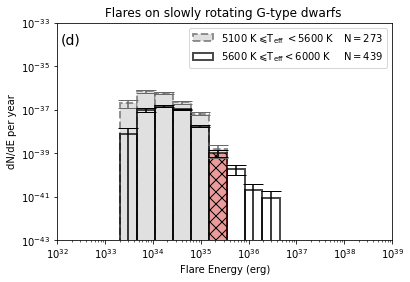} }} 
	\end{subfigure}
		\caption{The same as in Figures \ref{histogram-0-6} (a-d) and \ref{histogram-7-17} but for Q0-Q17 data. The flare frequency distribution in (b) for all G-type dwarfs and slowly rotating G-type dwarfs have power-law indices of 2.0 and 2.2, respectively. The red bin in (d) was used to calculate the occurrence frequency of super flares with energy of $2.26\times10^{35}$ erg.} 
		\label{histogram-0-17}
\end{figure}

Combining the two previous observations, we find that during 1424 days of continuous monitoring, we found 4637 super-flares on 1896 G-type dwarfs. Among them are 712 super-flares on 355 slowly rotating stars. We discovered 439 super-flares on 243 of the Sun-like stars. Our findings suggest that 397 stars out of 1896 G-type dwarfs have more than one super-flares, whereas 1499 stars have only one super-flare. A total of 55 out of 243 Sun-like stars have several super-flares, whereas 188 Sun-like stars have just one super-flare. Table \ref{tab:parameters} lists some super-flares stars with their parameters and the number of occurred super-flare $N_{\rm  f}$. The entire results can be found at \url{https://github.com/akthukair/AFD}. Figure \ref{histogram-0-17} similar to Figure \ref{histogram-7-17} shows four log-log scale histograms of the frequency distribution of super-flares as a function of flare energy in quarters 0 to 17. Figure \ref{histogram-0-17}(a) shows the distribution of the number of observed super-flares per observed flare peak amplitude. The amplitude of the 4,637 detected super-flares ranges between $7.5\times 10^{-4}$ and 35.6. Figure \ref{histogram-0-17}(b, c, and d) format is similar to Figure \ref{histogram-7-17}(b, c and d). Figure \ref{histogram-0-17}(b) compares the frequency distributions of super-flares on all G-type dwarfs (white-solid histogram) and slowly rotating G-type dwarfs (grey-dashed histogram). Super-flare frequency distributions on all G-type dwarfs have a power-law index of $\alpha \simeq 2.0 \ \pm 0.1$, while slowly rotating G-type dwarfs have a power-law index of $\alpha \simeq 2.2 \ \pm 0.3$. A comparison of the distribution of super-flares as a function of flare energy for all G-type dwarfs and slowly rotating G-type dwarfs, according to the effective temperature of the star are shown in Figures \ref{histogram-0-17}(c) and (d). In Figure \ref{histogram-0-17}(c), there are 897 super-flares on all cool G-type dwarfs (grey-dashed histogram) and 3740 super-flares on all hot G-type dwarfs (white-solid histogram), 273 super-flares on cool slowly rotating G-type dwarfs (grey-dashed histogram) and 439 super-flares on hot slowly rotating G-type dwarfs (white solid histogram) (d). The occurrence frequency rate of super-flares with energy equal to $2.26\times10^{35}$ erg is $2.29 \times 10^{-4}$ flares per year per star, corresponding to a super-flare occurring on a star once every 4360 years, as shown in the red bin in Figure \ref{histogram-0-17}(d).

\begin{table}
	\centering
	\caption{The parameters of super-flare stars (G-type) with the number of flares {\rm $N_{\rm  f}$}.}
	\label{tab:parameters}
	\begin{tabular}{cccccc}
		\hline
		\hline
		{\rm Kepler ID} & $T_{\rm  eff}$ & {\rm log g} & {\rm Radius} & $P_{\rm  rot}$ & {\rm $N_{\rm  f}$}\\
		& ($K$) & &($R_\odot$) & {\rm (day)} & \\
		\hline
		1718360 & 5908 & 4.44 & 1.00 & 2 & 5\\
		2302288 & 5904 & 4.45 & 1.01 & 9 & 1 \\
		2308761 & 5780 & 4.44 & 1.00 &	1 & 25 \\
		2445975 & 5780 & 4.44 & 1.00 & 1 & 9 \\
		2837133 & 5361 & 4.33 & 1.06 & 23 & 1 \\
		3852796 & 5650 & 4.37 & 0.98 & 11 & 2 \\
		3858729	& 5780 & 4.44 & 1.00 & 4 & 11 \\
		3869649 & 5525 & 4.46 & 0.90 & 1 & 13 \\
		4137840	& 5830 & 4.50 & 0.93 & 2 & 9 \\
		4346178	& 5392 & 4.31 & 1.09 & 34 & 3 \\
		4749912 & 5734 & 4.26 & 1.13 & 5 & 12 \\
		5105805	& 5849 & 4.52 & 0.92 & 13 & 2 \\
		5202404	& 5999 & 4.35 & 1.07 & 4 & 31 \\
		6223256 & 5951 & 4.30 & 1.11 & 2 & 38 \\
		6780893	& 5462 & 4.40 & 0.93 & 4 & 15 \\
		7958049 & 5802 & 4.38 & 0.94 & 2 & 97 \\
		8357647 & 5776 & 4.36 & 0.96 & 5 & 2 \\
		9390942 & 5999 & 4.49 & 0.96 & 8 & 1 \\
		9827094 & 5790 & 4.28 & 1.09 & 3 & 21 \\
		10518610 & 5530 & 4.43 & 1.02 & 18 & 1 \\
		10922936 & 5325 & 4.43 & 0.92 & 15 & 5 \\
		11809362 & 5216 & 4.45 & 0.90 & 15 & 1 \\
		11913716 & 5875 & 4.48 & 0.98 & 12 & 1 \\
		12266582 & 5638 & 4.39 & 0.93 & 6 & 4 \\
		12405306 &	5911 & 4.50 & 0.94 & 10 & 1 \\
		\hline
	\end{tabular}
		\begin{tablenotes}
		\small
		\item \textbf{Note:}\ The full version of the
		table is available 
		at \url{https://github.com/akthukair/AFD}
		\end{tablenotes}
\end{table}

\subsubsection{Comparison with \citet{Okamoto2021}}
\begin{figure}
	\centering
	\begin{subfigure}{\linewidth}
	\mbox{{\includegraphics[width=0.5\textwidth]{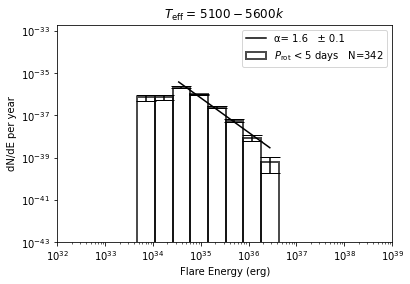}}\quad
		{\includegraphics[width=0.5\textwidth]{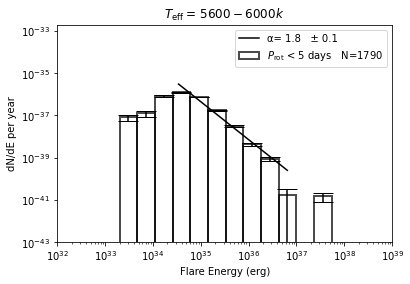} }}
	\mbox{{\includegraphics[width=0.5\textwidth]{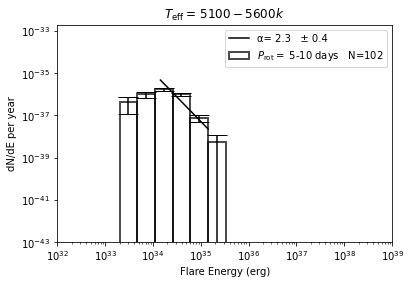}}\quad
		{\includegraphics[width=0.5\textwidth]{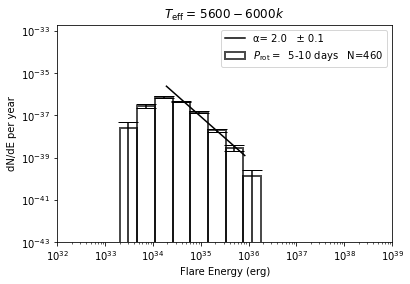} }}
	\mbox{{\includegraphics[width=0.5\textwidth]{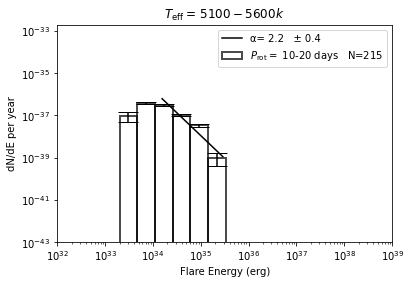}}\quad
		{\includegraphics[width=0.5\textwidth]{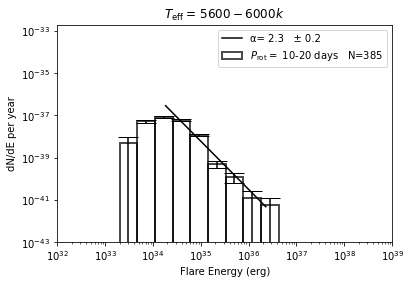} }}
	\mbox{{\includegraphics[width=0.5\textwidth]{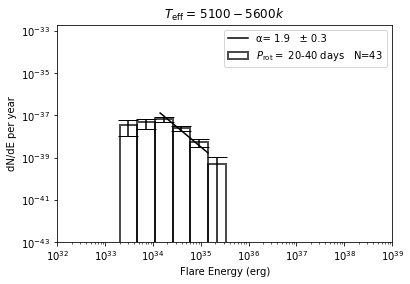}}\quad
		{\includegraphics[width=0.5\textwidth]{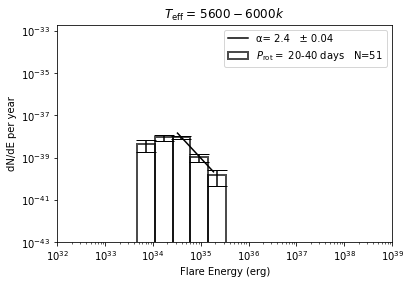} }} 
	\end{subfigure}
		\caption{The flare frequency distributions as a function of the flare energy ($dN/dE\propto E^{-\alpha }$) for different ranges of rotation periods ($P_{\rm rot}$: $<5$ days, 5-10 days, 10-20 days, 20-40 days) divided by the effective temperature $T_{\rm eff}$= 5100-5600 K on the left panel and $T_{\rm eff}$= 5600-6000 K on the right panel. $\alpha$ is the power law index, and N is number of super-flares for each $P_{\rm rot}$.\\} \label{Prot}
\end{figure}

In section \ref{subsubsection:0-6}, we compared our results with those of \citet{Shibayama_2013} since that section discusses the super-flares statistical analysis in (Q0-Q6) only as in \citet{Shibayama_2013} study. 
Here we compared our findings to \citet{Okamoto2021}, who used the complete Kepler primary mission data set and the Gaia Data Release 2 catalog to conduct the most recent statistical analyses of super-flares on solar-type (G-type main-sequence) stars.
\begin{enumerate}[label=(\roman*)]
	\item The sample size used to search for super-flare differs between the two studies. \citet{Okamoto2021} limit the study to 11,601 stars whose brightness variation amplitude and rotation period values were reported by \citet{McQuillan2014}. In this study, our sample size is more than two times larger than \citet{Okamoto2021} (25442 G-type main-sequence stars).
	\item We both used the same method used in previous studies \citep{Maehara2012, Maehara2015, Shibayama_2013} to detect super flares, except that \citet{Okamoto2021} improved the technique by developing an improved version of the flare-detection method, in which a high-pass filter was used to remove rotational variations caused by starspots. Furthermore, the sample biases on the frequency of super-flares were examined, taking both gyrochronology and the completeness of the flare detection into account. Our method does not include these improvements.
	\item We found 4637 super-flares on 1896 G-type dwarfs. The number of super-flares and super-flare stars is $\sim 2$ and $\sim 7$ times larger than the 2341 super-flares and 265 super-flare stars in \citet{Okamoto2021}, respectively. In addition, we found 439 superflares on 243 Sun-like stars based on \citet{Shibayama_2013} definition for the Sun-like stars ( $5600 K \leqslant T_{\rm  eff} < 6000 K$,  $\log \ g > 4.0$, and $P_{\rm rot}> 10$ days). If we consider the rotation period range in \citet{Okamoto2021} for the Sun-like stars ($P_{\rm rot}$ = 20 \textendash\, 40 days), the number of flares would be reduced to 51 super-flares on 38 Sun-like stars. This is approximately twice the number of super-flares and Sun-like stars that were discovered in \citet{Okamoto2021}, with 26 super-flares on 15 Sun-like stars.
	\item Our result includes 183  flare events on 41 G-type dwarfs, similar to those observed by \citet{Okamoto2021}. Table \ref{tab:common flare events} presents a comparison of these events between the two studies. Showing the parameters for each Kepler ID, the flare peak date and its energy. This result constitute 3.95\% and 7.82\% of the total flares detected in our work and \citet{Okamoto2021}, respectively. This low percentage in similar events is due to the different Kepler IDs in the samples used in the two studies. Since the effective temperature and stellar radius values in \citet{Okamoto2021} were taken from Gaia-DR2 in \citet{Berger2018}, resulting in different Kepler IDs for G-type main-sequence stars. We observed an increase in the flare energy estimate of our study compared to \citet{Okamoto2021}, with a correlation coefficient of 0.91 between the two energies.
	\item We calculated $\alpha$ index for the flare frequency distributions as a function of the flare energy ($dN/dE \propto E^{-\alpha}$) for different range of $P_{\rm rot}$ ($< 5$ days, 5-10 days, 10-20 days an 20-40 days) in two effective temperatures ranges (5100-5600 K and 5600-6000 K) as shown in Figure \ref{Prot} and compared our results with those of \citet{Okamoto2021} as shown in table \ref{tab:prot2}. The results are very similar, with $\alpha$ indexes around -2, confirming \citet{Okamoto2021} previous finding, which is also aligns with earlier research on super-flares on solar-type stars \citep{Shibata2013, Shibayama_2013, Maehara2015}. From this similarity, we believe that the effects applied in \citet{Okamoto2021} by applying a high-pass filter and examining the sample biases on super-flare occurrence rates considering gyrochronology and the completeness of flare detection have no significant affect on our final results.
	\item We used a different stellar parameters catalog (q1\_q17\_dr25\_ stellar catalog) however, we obtained similar results. Different stellar properties in the Gaia catalog, especially for radius, did not significantly affect our results, as shown in Table \ref{tab:common flare events}, as the correlation coefficient between energies in the two data sets $\simeq 0.91$.
\end{enumerate}

\begin{table*}[h!]
	\centering
	\tiny
	\caption{Comparison between the common flare events detected in this work and those observed by \citet{Okamoto2021}.}
	\label{tab:common flare events}
	\begin{adjustbox}{width=1\textwidth}
	\begin{tabular*}{\linewidth}{c|c|ccccc|ccccc}
		\hline
		\hline
		& &	\multicolumn{5}{c|}{This work} & \multicolumn{5}{c}{\citet{Okamoto2021}} \\
		{Kepler ID}&{Peak Date $^{\rm a}$}&$T_{\rm eff}$$^{\rm b}$&{Radius}$^{\rm b}$ &$P_{\rm rot}$&{Duration}&{Energy}& $T_{\rm eff}$$^{\rm c}$&{Radius}$^{\rm c}$&$P_{\rm rot}$$^{\rm d}$&{Duration}&{Energy}\\
		
		&(JD) &($\rm K$) &($\rm R_\odot$) &{\rm (days)}& {\rm (day)} & {\rm (erg)}& ($\rm K$) &($\rm R_\odot$)  &{\rm (days)}& {\rm (day)}& {\rm (erg)}\\
		\hline
		3217852&55648.38&5851&0.923&16.0&0.06&9.79$\times 10^{33}$&5852&1.137&17.04&0.06&6.60$\times 10^{33}$\\
		3217852&56021.66&5851&0.923&16.0&0.06&1.37$\times 10^{34}$&5852&1.137&17.04&0.06&7.30$\times 10^{33}$\\
		3217852&56050.95&5851&0.923&16.0&0.08&1.76$\times 10^{34}$&5852&1.137&17.04&0.08&1.40$\times 10^{34}$\\
		3217852&56084.25&5851&0.923&16.0&0.06&1.37$\times 10^{34}$&5852&1.137&17.04&0.08&1.70$\times 10^{34}$\\
		3217852&56160.70&5851&0.923&16.0&0.06&1.10$\times 10^{34}$&5852&1.137&17.04&0.06&1.10$\times 10^{34}$\\
		3853938&55197.55&5929&0.913&2.0&0.08&5.83$\times 10^{34}$&5928&1.418&2.22&0.1&9.60$\times 10^{34}$\\
		3853938&55341.92&5929&0.913&2.0&0.06&3.58$\times 10^{34}$&5928&1.418&2.22&0.1&6.90$\times 10^{34}$\\
		3853938&55689.27&5929&0.913&2.0&0.06&3.59$\times 10^{34}$&5928&1.418&2.22&0.06&4.80$\times 10^{34}$\\
		3869649&55004.30&5525&0.901&1.0&0.20&2.98$\times 10^{35}$&5524&0.859&1.92&0.21&1.90$\times 10^{35}$\\
		3869649&55061.78&5525&0.901&1.0&0.10&1.55$\times 10^{35}$&5524&0.859&1.92&0.08&5.70$\times 10^{34}$\\
		3869649&55202.44&5525&0.901&1.0&0.08&2.21$\times 10^{35}$&5524&0.859&1.92&0.19&9.70$\times 10^{34}$\\
		3869649&55243.90&5525&0.901&1.0&0.14&4.17$\times 10^{35}$&5524&0.859&1.92&0.21&2.70$\times 10^{35}$\\
		3869649&55268.36&5525&0.901&1.0&0.10&2.94$\times 10^{35}$&5524&0.859&1.92&0.21&2.50$\times 10^{35}$\\
		3869649&55348.09&5525&0.901&1.0&0.06&9.75$\times 10^{34}$&5524&0.859&1.92&0.21&7.10$\times 10^{34}$\\
		3869649&55656.76&5525&0.901&1.0&0.14&2.53$\times 10^{35}$&5524&0.859&1.92&0.23&2.20$\times 10^{35}$\\
		3869649&55680.83&5525&0.901&1.0&0.06&9.29$\times 10^{34}$&5524&0.859&1.92&0.1&4.00$\times 10^{34}$\\
		3869649&55684.08&5525&0.901&1.0&0.10&1.87$\times 10^{35}$&5524&0.859&1.92&0.21&1.60$\times 10^{35}$\\
		3869649&55732.00&5525&0.901&1.0&0.06&9.95$\times 10^{34}$&5524&0.859&1.92&0.17&6.20$\times 10^{34}$\\
		3869649&56342.90&5525&0.901&1.0&0.06&7.01$\times 10^{34}$&5524&0.859&1.92&0.12&4.10$\times 10^{34}$\\
		3869649&56347.33&5525&0.901&1.0&0.10&1.29$\times 10^{35}$&5524&0.859&1.92&0.21&8.40$\times 10^{34}$\\
		4276035&55975.58&5210&0.979&23.77 $^{\rm d}$&0.06&2.35$\times 10^{34}$&5212&0.755&23.77&0.08&1.00$\times 10^{34}$\\
		4276035&56260.53&5210&0.979&23.77 $^{\rm d}$&0.06&3.20$\times 10^{34}$&5212&0.755&23.77&0.1&1.30$\times 10^{34}$\\
		4749912&54975.49&5734&1.127&5.0&0.08&6.57$\times 10^{34}$&5736&0.845&5.07&0.1&2.30$\times 10^{34}$\\
		4749912&55020.65&5734&1.127&5.0&0.06&3.81$\times 10^{34}$&5736&0.845&5.07&0.06&1.00$\times 10^{34}$\\
		4749912&55064.68&5734&1.127&5.0&0.06&3.81$\times 10^{34}$&5736&0.845&5.07&0.08&1.50$\times 10^{34}$\\
		4749912&55100.11&5734&1.127&5.0&0.06&3.39$\times 10^{34}$&5736&0.845&5.07&0.1&1.30$\times 10^{34}$\\
		4749912&55126.31&5734&1.127&5.0&0.06&4.94$\times 10^{34}$&5736&0.845&5.07&0.08&8.40$\times 10^{33}$\\
		4749912&55244.51&5734&1.127&5.0&0.06&1.00$\times 10^{35}$&5736&0.845&5.07&0.15&1.80$\times 10^{34}$\\
		4749912&55261.02&5734&1.127&5.0&0.10&1.26$\times 10^{35}$&5736&0.845&5.07&0.19&5.90$\times 10^{34}$\\
		4749912&55857.97&5734&1.127&5.0&0.08&7.36$\times 10^{34}$&5736&0.845&5.07&0.12&1.80$\times 10^{34}$\\
		4749912&55886.52&5734&1.127&5.0&0.08&7.36$\times 10^{34}$&5736&0.845&5.07&0.15&2.90$\times 10^{34}$\\
		4749912&55901.41&5734&1.127&5.0&0.10&9.58$\times 10^{34}$&5736&0.845&5.07&0.12&3.90$\times 10^{34}$\\
		4749912&56295.82&5734&1.127&5.0&0.08&7.38$\times 10^{34}$&5736&0.845&5.07&0.1&2.50$\times 10^{34}$\\
		5953631&55331.77&5911&0.997&14.0&0.06&6.41$\times 10^{34}$&5911&0.973&14.78&0.08&3.00$\times 10^{34}$\\
		6110415&55386.47&5857&1.045&7.0&0.06&1.86$\times 10^{34}$&5857&0.841&6.90&0.08&7.50$\times 10^{33}$\\
		6110415&55664.06&5857&1.045&7.0&0.06&3.01$\times 10^{34}$&5857&0.841&6.90&0.06&5.30$\times 10^{33}$\\
		6110415&55694.58&5857&1.045&7.0&0.08&4.41$\times 10^{34}$&5857&0.841&6.90&0.19&3.10$\times 10^{34}$\\
		6110415&56011.02&5857&1.045&7.0&0.06&1.54$\times 10^{34}$&5857&0.841&6.90&0.1&9.30$\times 10^{33}$\\
		6110415&56113.37&5857&1.045&7.0&0.08&3.17$\times 10^{34}$&5857&0.841&6.90&0.15&1.80$\times 10^{34}$\\
		\hline
	\end{tabular*}
	\end{adjustbox}
	\begin{tablenotes}
		\small
		\item \textbf{Note:}\ The full version of the table is available at \url{https://github.com/akthukair/AFD}\\
		\ $^{\rm a}$ Flare peak in Julian Date (JD).\\
		\ $^{\rm b}$ The effective temperature and stellar radius values in this work are derived from Data Release 25 (DR 25).\\
		\ $^{\rm c}$ The effective temperature and stellar radius values in \citet{Okamoto2021} are derived from Gaia-DR2 in \citet{Berger2018}.\\
		\ $^{\rm d}$ Rotation period from \citet{McQuillan2014}.
	\end{tablenotes}	
\end{table*}

\begin{table*}
	\centering
	\caption{Comparison of the $\alpha$ index for the flare frequency distributions as a function of the flare energy ($dN/dE \propto E^{-\alpha}$) for each $P_{rot}$ range.}
	\label{tab:prot2}
	\begin{tabular*}{\textwidth}{ @{\extracolsep{\fill}}c|cc|cc}
		\hline
		\hline
		\multirow{4}{*}\centering{} &		
		\multicolumn{2}{c|}{{$T_{\rm eff}=5100-5600$}} & 
		\multicolumn{2}{c}{{$T_{\rm eff}=5600-6000$}} \\
		& {This work} & {\citet{Okamoto2021}} & {This work} & {\citet{Okamoto2021}}\\
		\hline
		$P_{rot}<5$ days& $-1.6 \pm0.1$ & $-1.5 \pm0.1$ & $-1.8 \pm0.1$ & $-1.8 \pm0.1$ \\
		$P_{rot}=5-10$ days& $-2.3 \pm0.4$ & $-1.9 \pm0.2$ & $-2.0 \pm0.1$  & $-2.1 \pm0.1$ \\
		$P_{rot}=10-20$ days& $-2.2 \pm0.4$ & $-2.2 \pm0.2$ & $-2.3 \pm0.2$  & $-2.2 \pm0.1$\\
		$P_{rot}=20-40$ days& $-1.9 \pm0.3$ & $-2.1 \pm0.2$ & $-2.4 \pm0.04$  & $-2.7 \pm0.1$ \\	
		\hline		
	\end{tabular*}
\end{table*} 

\subsection{Super Flares on Other Spectral Type Stars}
We detected a total of 11438 super-flares on 1740 stars of other spectral types during 1424 days of continuous observation of dwarfs of 2653 A-type, 10898 F-type, 10307 K-type and 2222 M-type. The entire results can be found at \url{https://github.com/akthukair/AFD}.

For A-type dwarfs, we found 321 super-flares on 136 stars; 44 of these stars show more than one super-flare and 92 stars show only one super-flare. The duration of these super-flares can be up to 0.27 days, their flare maximum amplitude is 0.47. The largest flare energy we have determined is $8.91\times10^{37}\rm erg$, with an amplitude of 0.28, and it lasted for 0.16 days.

As for F-type dwarfs, we discovered 1125 super-flares on 522 stars; 106 of these stars exhibit several super-flares, whereas 416 stars exhibit only one. These super-flares duration can be up to 0.57 days, and their flare amplitude can reach 0.08. Our calculations show that the highest flare energy we have found was $2.35 \times 10^{36}\rm erg$, had an amplitude of 0.036, and lasted for 0.2 days.

Moving on to K-type dwarfs, we detected 4538 super-flares on 770 stars; 304 show several super-flares, while 466 stars show just one. The duration of these super-flares can be up to 0.3 days, and their amplitude can reach 1.04. The largest flare energy we have measured is $2.82\times10^{36}\rm erg$, with an amplitude of 0.057 and a duration of 0.28 days.

For M-type dwarfs, we found 5445 super-flares on 312 stars; 256 of which have several super-flares and 56 have only one. These super-flares duration can be up to 0.82 days, and their flare amplitude varies between 0.002 to 15.1. The highest flare energy we have recorded was $1.59\times10^{35}\rm erg$,  has an amplitude of 15.1 and a duration of 0.18 days.

\begin{figure}
	\centering
	\begin{subfigure}{\linewidth}
	\mbox{{\includegraphics[width=0.5\textwidth]{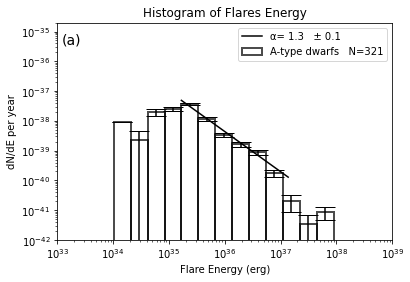}}\quad
		{\includegraphics[width=0.5\textwidth]{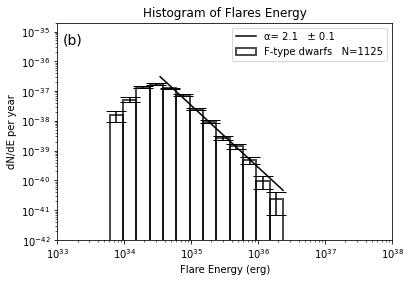}}}
	\mbox{{\includegraphics[width=0.5\textwidth]{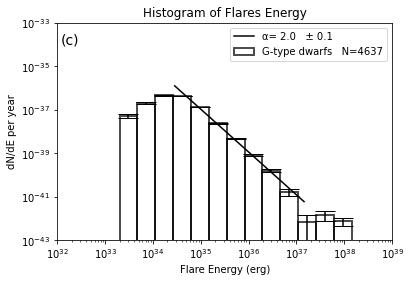}}\quad
		{\includegraphics[width=0.5\textwidth]{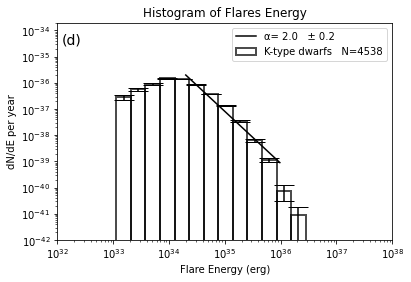}}}
	\mbox{{\includegraphics[width=0.5\textwidth]{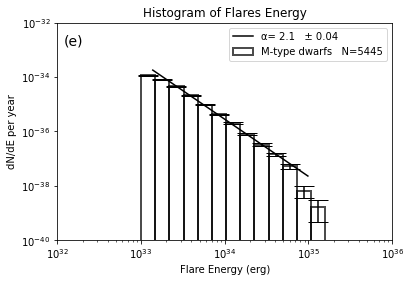} }}
	\end{subfigure}
		\caption{Log-log scale histograms showing the frequency distribution of super-flares organised by the stellar spectral type, with the $\alpha$ index of the power-law relation ($dN/dE \propto E^{-\alpha}$) and the number of super-flares for each spectral type indicated. We used approximately the same energy range for the fitting in all panels and ignored the bins towards the end of the histogram with a limited number of flares.\\} \label{spectral type sf}	
\end{figure}

Figure \ref{spectral type sf} shows the flares frequency distribution on different spectral types, which follows a power-law relation ($dN/dE \propto E^{-\alpha}$). We used roughly the same energy range to obtain the best fit for $\alpha$ index in order to compare between the results. In order to maintain accurate statistics, we disregarded bins at the end of the histogram that contain a small number of flares (five or less).

\citet{Yang2019} reported a total of 162,262 flare events on 3420 flaring stars. They found that the flare frequency distribution from F-type stars to M-type stars have an index $\alpha \sim$ 2, while $\alpha \sim$ 1 for the flares frequency distribution of A-type stars. In our study, we found that the $\alpha$ index is about 2 for spectral types from F-type to M-type stars. In particular $\alpha \simeq 2.1\ \pm0.1$ for F-type stars, $\alpha \simeq 2.0\ \pm0.1$ for G-type stars, $\alpha \simeq 2.0\ \pm0.2$ for K-type stars and $\alpha \simeq 2.1\ \pm0.04$ for M-type stars. These results are consistent with the previous results of \citet{Yang2019}.
\citet{Shibata2013} found that for the Sun-like stars, the flares frequency distribution of nanoflares, microflares, solar flares and superflares, follow a power-law relation with $\alpha \simeq 1.8$. It means that the underlaying mechanism which is generating these flares
by magnetic reconnection occurs in similar
physical conditions \citep{Yang2019, Cliver2022}. It is thought that the same fundamental process that causes solar flares also causes flares in late-type stars (F-M).  A convective envelope is necessary for the dynamo that generates the magnetic fields in these stars. For magnetic fields to become strong enough to rise and emerge into the stellar atmosphere, which results in the generation of flares, this convective envelope must be sufficiently deep \citet{Pedersen2017}. The similarity of $\alpha$ index indicates that the process of producing flares in various spectral types from F-type to M-type stars is similar and is based on magnetic reconnection \citet{Yang2019}.

However, the situation is different for A-type stars, as the $\alpha$ index for the flare frequency distribution differs from the rest and is
$\alpha \simeq 1.3\ \pm0.1$. It is widely expected that these stars cannot have flares. For stars to produce flares, they must have a deep outer convection zone, powerful large-scale magnetic fields, or powerful winds produced by the radiation. Normal A-type stars are devoid of these characteristics. Hence they should not flare \citet{Pedersen2017}. Moreover, According to stellar evolution theories, A-type stars are unlikely to flare because of their weak magnetic fields, and their thin or nonexistent surface convection zone, preventing a magnetic dynamo from operating \citet{Van-Doorsselaere2017}. However, through a visual inspection of the light curves, \citep{Balona2012, Balona2013} found flare events in the light curves of 33 A-type stars. To study the origin of these flares, \citet{Pedersen2017} reported a new, detailed analysis of these 33 A-type stars and verified the existence of flares in 27 of them. \citet{Balona2015} observed 1833 and 424 A-type stars in long cadence and 
short cadence modes, respectively, during Q0-Q12  using visual inspection. In the long cadence and 
short cadence modes, 51 and 10 A-type stars, respectively, show evidence of flare activity. 
Moreover, 24 new A-type stars with flaring activity have been discovered by \citet{Van-Doorsselaere2017}. 

Table \ref{tab:excluded flares} shows the number of super-flares for each spectral type. Where $N_{\rm f\ (candidates)}$ is the number of flares candidates captured by the code when the flux difference exceeded the threshold limit, $N_{\rm f\  (conditions)}$ is the number of flares that met all the conditions, $N_{\rm f\ (check)}$ is the number of flares we verified using visual inspection of the light curves, and $N_{\rm f\ (exclude)}$ is the number of flares that we excluded due to their irregular, chaotic shape after the visual inspection of the light curves. By analyzing this table, we notice a large fraction of excluded flares are in the A-type stars compared to the M-type stars. This can be explained by the different flare conditions in the stars. One difference can be that the sizes of A-type star-spots are much larger than that of M-type, with different magnetic reconnection conditions,
resulting in the flares with irregular, chaotic shapes of the light curves in A-type stars. This is possibly why A-type stars have a different $\alpha$ index. 
Table \ref{tab:flare incidence} shows the incidence of flares for each stellar spectral type. Since our study considers main-sequence stars only, the number of stars is limited to a specific radius range for each spectral type, and therefore the number of stars in this study is much smaller than in other studies such as \citet{Yang2019}. Thus the flare incidence appears in larger numbers than in those studies. The flare incidence gradually increases from F-type to M-type stars from 4.79~\% to 14.04~\% because of the increase of the convection zone depth \citet{Yang2019}. However, the incidence of the flares in A-type stars 5.13~\% is higher than in F-type stars 4.79~\%, which contradicts the theoretical expectation. As from an A-type star to an F-type star, a star's outer layer changes from the radiative envelope to the convective envelope, allowing F-type stars to operate a solar-like dynamo. In contrast, A-type stars struggle to create and maintain a magnetic field \citet{Yang2019}. These results are also consistent with \citep{Balona2015,Van-Doorsselaere2017,Yang2019}, who analyzed stars of any size, not just those on the main-sequence discussed here. The deviation of the $\alpha$ index in A-type stars from the rest of stellar types, 
the rise of flare incidence rate compared to F-type stars, and the high percentage of excluded flares compared to the rest of the spectral types are all indications that A-type stars may generate flares in a different,
peculiar manner.

\begin{table*}
	\caption{The ratio rate of excluded flares for each spectral type.}
	\label{tab:excluded flares}
	\begin{tabular}{cccccc}
		\hline
		\hline
		{\rm Class} &$N_{\rm  f\ (candidates)}$ & $N_{\rm  f\  (conditions)}$ & $N_{\rm  f\ (check)}$ & $N_{\rm  f\ (exclude)}$ & $N_{\rm  f\ (exclude)}/ N_{\rm  f\ (conditions)}$ \\
		\hline
		\rm A & 38876 & 953 & 321 & 632 & 66.32\ \% \\
		\rm F & 187877 & 2484 & 1125 &1359 & 54.71\ \% \\
		\rm G & 481005 & 6791 & 4637 & 2154 & 31.72\ \% \\
		\rm K & 207205 & 6754 & 4538 & 2216 & 32.81\ \% \\
		\rm M & 15165 & 6808 & 5445 & 1363 & 20.02\ \% \\		
		\hline		
	\end{tabular}
	\begin{tablenotes}
		\small
		\item \textbf{Note:}\ $N_{\rm f\ (candidates)}$ represents the number of flare candidates captured by our Python code when the flux difference exceeds the threshold limit. $N_{\rm f\  (conditions)}$ is the number of flares that satisfied all conditions, $N_{\rm f\ (check)}$ is the number of flares that were confirmed by visual examination of the light curves. Following a visual examination of the light curves, $N_{\rm f\ (exclude)}$ is the number of flares that were excluded due to their irregular, chaotic shape.\\
	\end{tablenotes}	
\end{table*}

\begin{table}
	\centering
	\caption{The number of stars $N_{\rm star}$, flare stars $N_{\rm fstar}$ and flare incidence for each spectral type.}
	\label{tab:flare incidence}
	\begin{tabular}{cccccc}
		\hline
		\hline
		{\rm Class} & $T_{\rm eff}$ & {\rm Radius} & $N_{\rm  star}$ & $N_{\rm  fstar}$&{\rm incidence} \\
		& $(\rm K)$ &($\rm R_\odot$) & & &\\
		\hline
		\rm A & 7500 - 10000 & 1.4 - 1.8 &2653 & 136 & 5.13\ \%\\
		\rm F & 6000 - 7500 & 1.15 - 1.4 & 10898 & 522 & 4.79\ \%\\
		\rm G & 5100 - 6000 & 0.9 - 1.15 &25442 & 1896 & 7.45\ \%\\
		\rm K & 3700 - 5200 & 0.7 - 0.96 &10307 & 770 & 7.47\ \%\\
		\rm M & 2400 - 3700 & $\leqslant$ 0.7 &2222 & 312 & 14.04\ \%\\		
		\hline
	\end{tabular}	
\end{table}

\section{Conclusion}\label{section:conclusion}
Using bespoke Python script written by us, we performed an automated search for super-flares on main-sequence stars of types A, F, G, K, and M in all of Kepler's long-cadence data of DR 25 from Q0 to Q17, using \citep{Maehara2012,Shibayama_2013} technique. We used the Harvard Spectral classification to determine each target's spectral type based on its effective temperature and radius. For A, F, G, K and M type stars, we studied a total of 2653, 10898, 25442, 10307, and 2222 main-sequence stars, respectively. As a result, we detected 4637 super-flares on 1896 G-type dwarfs during 1424 days of continuous observation by Kepler. Using these new data, we studied and compared the statistical properties of the occurrence frequency rate of super-flares using three distinct datasets, namely, Q0-Q6, Q6-Q17, and Q0-Q17. According to our estimates for the Q0-Q17 dataset, a super-flare on G-type dwarfs with an energy of $10^{35}$ erg occurs on a star once every 4360 years. By comparing three data sets (0-6, 7-17 and 0-17), the statistics of flare look very similar, which is indicative of the fact that flare occurring statistic does not change on the time scale of 17 quarters. In addition, we compared our results with those of \citet{Okamoto2021} regarding the flare frequency distribution FFD and the $\alpha$ index for the power-law relation. The similarity of the results supports the consistency with \citet{Okamoto2021} prior research. This suggests that the high-pass filter and analysis of sample biases on super-flare occurrence rates that \citet{Okamoto2021} performed while considering gyrochronology and the completeness of the flare detection had no appreciable impact on our final results, which is one of the motivations for our study. Also, we compared 183 similar flare events to \citet{Okamoto2021} on a case-by-case basis, as in Table 7. We noted an overestimation of the flare energy for our study compared to \citet{Okamoto2021}. Despite using a different catalog of stellar parameters, available to us, we found that the results are consistent with \citet{Okamoto2021}. The different stellar parameters in the Gaia catalog had no significant effect on our findings as illustrated in Table \ref{tab:common flare events}. Moreover, we detected a total of 321, 1125, 4538 and 5445 super-flares on dwarfs of 136, 522, 770 and 312 A, F, K and M type stars respectively, during 1424 days of continuous observation. We determined the distributions of super-flare occurrence rates as a function of super-flare energy. We found that for all spectral types of stars, from F-type to M-type, the flare frequency distribution as a function of flare energy follows a power-law relation with $dN/dE \propto E^{-\alpha}$ where $\alpha \simeq$ 2.0 to 2.1. This demonstrates that the power-law index $\alpha$ is similar to those of solar flares ($\sim 2$) \citet{Shibayama_2013}. The power-law index values' similarity suggests that similar physical conditions produce the flares, by the mechanism which is thought to be magnetic reconnection. In contrast, the obtained value of $\alpha$ index 1.3 of the flare frequency distribution for A-type stars indicates that their flare conditions are distinct from those of the other stellar types. We observed a general rise in the flare incidence rate from 4.79~\% and 14.04~\% for F-type to M-type stars. However, the flare incidence rate is higher in A-type stars 5.13~\% than in F-type stars 4.79~\%, contrary to the theoretical expectation \citet{Yang2019}. These results are similar to those found by \citep{Balona2015,Van-Doorsselaere2017,Yang2019}.

In general, flares with smaller amplitude tend to be more challenging to
detect. Therefore, the detection completeness of flares with smaller
amplitude is lower than that of larger ones, and the actual frequency of
flares with smaller amplitude would be higher than the observed one.
In addition, as discussed in \citet{Okamoto2021}, the flare-detection
threshold proposed by \citep{Maehara2012,Shibayama_2013} depends on the
rotation period and the amplitude of rotational modulations.
In relation to the evaluation and discussion on the flare-detection
completeness, which has a significant impact on the power-law slope of
the flare-frequency distribution (FFD), we would like to remark that
since the power law slope for G type-stars is similar to that of \citet{Shibayama_2013} and for A-type stars that of \citet{Yang2019}, we believe that
a lack of dedicated analysis of flare-detection completeness has no
significant impact on our results.

In relation to the contamination of flares on close neighboring stars we
would like to remark that since the spatial resolution of the Kepler
space telescope is not high ($\sim 4$ arcsec/pixel), some stars have close
neighbors within the photometric aperture. As pointed out by \citep{Maehara2012, Pedersen2017}, flares on close neighboring stars
can produce false flare signals in the target light curve. These false
flares may potentially have an impact on the flare frequency. Although we
excluded the stars having neighboring stars within 12-arcsec, we believe that this is sufficient despite the fact that not all faint
stars are cataloged. Again our belief is backed up by the fact that the
power law slope for G type-stars is similar to that of \citet{Shibayama_2013}
and for A-type stars that of \citet{Yang2019}.

In relation to how we determine the flare energy range used for fitting
the observed FFD to the power-law function, we would like to acknowledge
the fact that the number of flares in the lower energy bins would be
much smaller than the actual flare frequency. Therefore, one should
choose the energy range in which the detection completeness issue can be
negligible. In our case, the choice which leftmost energy range bins to
ignore was done intuitively by eye, similar to \citet{Yang2019}.

\section*{Acknowledgements}
Some of the data presented in this paper were obtained from the Mikulski Archive for Space Telescopes (MAST). STScI is operated by the Association of Universities for Research in Astronomy, Inc., under NASA contract NAS5-26555. Support for MAST for non-HST data is provided by the NASA Office of Space Science via grant NNX13AC07G and by other grants and contracts.\\
Authors would like to thank Deborah Kenny of STScI for kind assistance in obtaining the data, Cozmin Timis and Alex Owen of Queen Mary University of London for the assistance in data handling at the Astronomy Unit.\\
A. K. Althukair wishes to thank Princess Nourah Bint Abdulrahman University, Riyadh, Saudi Arabia and  
Royal Embassy of Saudi Arabia Cultural Bureau in London, UK
for the financial support of her PhD scholarship, held at
Queen Mary University of London.\\

%%%%%%%%%%%%%%%%%%%%%%%%%%%%%%%%%%%%%%%%%%%%%%%%%%
\section*{Data Availability}
All data used in this study was generated by our bespoke Python script that can be found at \url{https://github.com/akthukair/AFD} under the filename AFD.py and other files in the same GitHub repository.
The data underlying this article were accessed from Mikulski Archive for Space Telescopes (MAST) \url{https://mast.stsci.edu/portal/Mashup/Clients/Mast/Portal.html}. The long-cadence Kepler light curves analyzed in this paper can be accessed via MAST \citet{https://doi.org/10.17909/t9488n}, {\url{https://doi.org/10.17909/t9488n}}. The Kepler Stellar parameters table for all targets can be found at \citet{https://doi.org/10.26133/nea6}. The derived data generated in this research will be shared on reasonable request to the corresponding author.

\bibliography{RAA-2023-0150.R1}{}
\bibliographystyle{aasjournal}

\end{document}